\documentclass[11pt]{article}

\pdfoutput=1 

\usepackage{jheppub} 
\usepackage[T1]{fontenc} 

\usepackage{subfig,amsmath,amsfonts,amssymb}
\usepackage{todonotes}
\usepackage{lmodern}
\usepackage{tikz}
\usetikzlibrary{decorations.pathmorphing}
\usepackage{caption}
\usepackage{subfig}
\usepackage{graphicx}
\usepackage{verbatim}
\usepackage{cleveref}
\usepackage{braket}
\tikzset{snake it/.style={decorate, decoration=snake}}

\title{Liouville conformal blocks and Stokes phenomena}
\author[\ast]{Xia Gu,}
\author[\ast,\dag]{Babak Haghighat,}
\affiliation[\ast]{Yau Mathematical Sciences Center, Tsinghua University, Beijing, 100084, China}
\affiliation[\dag]{Yanqi Lake Beijing Institute of Mathematical Sciences and Applications (BIMSA), Huairou District, Beijing 101408, P. R. China}
\emailAdd{gux19@mails.tsinghua.edu.cn}
\emailAdd{babakhaghighat@tsinghua.edu.cn}

\abstract{In this work we derive braid group representations and Stokes matrices for Liouville conformal blocks with one irregular operator. By employing the Coulomb gas formalism, the corresponding conformal blocks can be interpreted as wavefunctions of a Landau-Ginzburg model specified by a superpotential $\mathcal{W}$. Alternatively, these can also be viewed as wavefunctions of a 3d TQFT on a 3-ball with boundary a 2-sphere on which the operator insertions represent Anyons whose fusion rules describe novel topological phases of matter.}

\date{}

\begin{document}
\maketitle

\section{Introduction}
Conformal field theory in two dimensions can be exactly solved due to an infinite-dimensional group of conformal transformations \cite{Belavin:1984vu}. The correlation functions, being single-valued, are organized in terms of products of \textit{conformal blocks} which are themselves not single-valued and transform non-trivially upon braiding of operators \cite{Moore-Seiberg}. Computation of such conformal blocks was soon found to be facilitated through the so-called \textit{Coulomb gas} formalism of Dotsenko and Fateev \cite{Dotsenko:1984nm}. Within this formalism, one views the individual operator insertions as charged particles subject to a two-dimensional Coulomb potential whose partition function gives rise to the conformal blocks. This viewpoint had a major impact on numerous constructions in conformal field theory, and in particular led to the discovery that degenerate Virasoro conformal blocks form braid group representations associated to the Jones polynomial \cite{cmp/1104201923,Varchenko}. In this context, we would like to note that integration cycles used to compute conformal blocks in the Coulomb gas formalism were usually compact and all constructions referred to so far were using such compact cycles. 

One major change of viewpoint arises by viewing the 2d CFT as a boundary of a 3d TQFT \cite{Witten:1988hf}. In this context, the 2d conformal blocks are interpreted as boundary wavefunctions of three-dimensional Chern-Simons theory. In cases where the Chern-Simons theory has a compact gauge group, one can recover a large class of rational CFTs including the WZW models and coset models \cite{Witten:1988hf,Witten:1991mm}. In such cases, the number of conformal blocks is always finite as the boundary Hilbert space is finite-dimensional. When the gauge group of the bulk 3d theory is non-compact, the boundary Hilbert space becomes infinite-dimensional. CFTs with an infinite number of conformal blocks are non-rational and depend on a further complex parameter, denoted here by $b$. Examples are Liouville and Toda theory corresponding to bulk 3d theories with $SL(2,\mathbb{C})$ and $SL(N,\mathbb{C})$ ($N > 2$) gauge groups. Such Chern-Simons theories admit novel and interesting phenomena including integration over so-called \textit{Lefschetz Thimbles} \cite{Witten:2010cx}. One can still obtain rational CFTs from these by taking the parameter $b$ to have certain discrete values. In particular, setting $b^2 = -\frac{m}{n}$, with $m$, $n$ co-prime positive integers, ensures that the resulting theory is a minimal model \cite{VafaFQHE}. An important consequence of this is the interpretation of the corresponding conformal blocks as Fractional Quantum Hall wavefunctions as advocated in \cite{VafaFQHE,Bergamin:2019dhg}. Taking a yet broader viewpoint, one could ask about the entirety of topological phases which can be obtained employing this formalism. 

In the current work, we would like to take the above viewpoint while following the approach of \cite{Gaiotto:2011nm} (see also \cite{Galakhov:2016cji,Galakhov:2017pod}) to construct conformal blocks which are interpreted as wavefunctions of a 2d Landau-Ginzburg theory specified by a superpotential $\mathcal{W}$. One novelty of this approach is that the conformal blocks are constructed by integrating over non-compact cycles and moreover irregular singularities are considered. Irregular singularities lead to Stokes phenomena where asymptotic expansions are only valid in certain wedges in the complex plane and have to be multiplied by Stokes matrices upon crossing co-dimension one walls. This complicates the analysis of monodromy and braiding phenomena but provides also a more interesting and novel perspective towards braiding. From the point of view of conformal field theory and Liouville theory, irregular operators have been studied in \cite{Gaiotto:2012sf} (see also \cite{Bonelli:2022ten} for a more recent treatment connected to our current work) where it was argued that one can think of them as certain collisions of two regular punctures. We will be studying 2-point and 3-point correlators with irregular punctures and analyze their monodromy and braid group representations. One important application we have in mind is with regard to topological quantum computation and the question of universality \cite {Freedman:modular} for this new class of representations. Analogous to the case of Ising- and Fibonacci-Anyons where an analysis of the monodromy of conformal blocks leads to $F$- and $R$-matrices \cite{Gu:2021utd} and thus fixes the fusion category, we find that a similar analysis including Stokes phenomena leads to braiding matrices $B_{12}$ and $B_{23}$. However, the situation is more intricate here and it does not seem that such matrices can be expressed in terms of $F$- and $R$-matrices in the usual way. It appears that some of our Anyons, namely those corresponding to irregular singularities, belong to nonsimple objects in the fusion category \cite{Fusion:2005} and the question arises as how to extend the standard notion of a fusion category to accommodate them. We will not answer this question in the current work, but hope to have made some progress towards this direction.

The organization of the present paper is as follows. In Section \ref{sec:asp} we review basics of Liouville field theory and the fusion rules of degenerate primary fields. In Section \ref{sec:thimbles} we describe the Thimble representation of conformal blocks using steepest descent paths corresponding to a superpotential $\mathcal{W}$. We then illustrate the occurrence of Stokes phenomena in this context which can be seen as a jump of integration cycles when crossing co-dimension one lines. In Section \ref{sec:monodromy} we describe the monodromy of conformal blocks in terms of formal monodromy and Stokes matrices. As an example we present the case of the Modified Bessel functions which correspond to certain 2-point correlators together with an irregular singularity at infinity. We then move on to our main example, namely that of 3-point functions of degenerate fields with an irregular singularity at infinity. We solve for the formal monodromy and the Stokes matrices and finally derive a representation of the braiding matrices in Section \ref{sec:braiding}. Finally, in Section \ref{sec:conclusions} we present our conclusions.

\section{Aspects of Liouville field theory}
\label{sec:asp}
Here we briefly review the basic facts about the Liouville field theory, which would be useful in the computation of the conformal blocks. The general references are \cite{Teschner_2001,https://doi.org/10.48550/arxiv.hep-th/9304011}.
Liouville field theory is a $2d$ CFT, parametrized by the ``coupling constant'' $b$. Classically, its action is 
\begin{equation}
S=\int dtd\sigma \left(\frac{1}{16\pi}((\partial_t\varphi)^2-(\partial_\sigma\varphi)^2)-\mu e^\varphi\right).
\end{equation}
The corresponding quantum theory depends on the Planck constant $\hbar=b^2$. The Liouville field $\phi$ correspond to $\varphi$ by $\phi\underset{b\rightarrow 0}{\sim}\frac{1}{2b}\varphi$.

The primaries in this theory are $V_\alpha=e^{2\alpha \phi}$ labeled by $\alpha\in\mathbf{C}$. Sometimes $\alpha$ is called the momentum of the primary.  The central charge $c$ of the theory and the conformal dimension $\Delta(\alpha)$ of primary operators $V_{\alpha}$ are determined by
\begin{gather}
	Q=b+\frac1b,\; c=1+6Q^2,\; \Delta_\alpha=\alpha(Q-\alpha).
\end{gather}

In a sense, we can regard Liouville field theory as a generalization of the Coulomb gas representation of the unitary minimal models. To see this, we need to introduce some notation which connects to minimal models. Note that for real $b$, we have central charge $c\geq 25$. So for the case related to minimal models, $b$ is some pure imaginary number.

Let's define 
\begin{equation}
\beta^2=-(Q-2\alpha)^2.
\end{equation}
Then we can re-parameterize $\Delta_\alpha$ in terms of $\beta$ as 
\begin{equation}
\begin{gathered}
    h_0=\frac{1}{24}(c-1),\\
    \Delta_\alpha(\beta)=\alpha(Q-\alpha)=h_0+\frac14\beta^2.
\end{gathered}
\end{equation}
We also use $\Phi_\beta$ to denote the fields with conformal dimension $h=\Delta_\alpha(\beta)$ along with $V_\alpha$. 

There is a special set of fields $\Psi_{r,s},\;r,s\in \mathbf{N}^+$ that have simple fusion rules with other primary fields. They are called degenerate fields and they have the following conformal dimension: 
\begin{equation}
    \begin{gathered}
    h_{r,s}=h_0+\frac{1}{4}(r a_++sa_-)^2,\\
 a_{\pm}=\frac{\sqrt{1-c}\pm\sqrt{25-c}}{\sqrt{24}}.
    \end{gathered}
\end{equation}
In Liouville theory these degenerate fields can be constructed using the Liouville field $\phi$ by \cite{VafaFQHE}:
\begin{equation}
\label{eq:degenerate fields}
     \Psi_{r,s}=\exp\left([(1-r)b+\frac{1-s}{b}]\phi\right),
\end{equation}
with $a_+=i b,\; a_-=i b^{-1}$.

 $\Psi_{r,s}$ can realize the fusion rules of $(p',p)$ minimal models, as long as we adjust the value of $b$ to match the central charge of the corresponding minimal model. Namely, we should set $Q^2=(b+\frac{1}{b})^2=\frac{-(p-p')^2}{pp'}$. The fusion algebra is closed among fields with $1\leq r< p',\; 1\leq s<p$ \cite{DiFrancesco:639405}:
\begin{equation}    \Psi_{r_1,s_1}\times \Psi_{r_2,s_2}=\sum_{k=1+|r_1-r_2|}^{k_{max}}\sum_{l=1+|s_1-s_2|}^{l_{max}}\Psi_{k,l}\;,
\end{equation}
wherein 
\begin{equation}
\begin{aligned}
        k_{max}&=\text{min}(r_1+r_2-1,2p'-1-r_1-r_2),\\
        l_{max}&=\text{min}(s_1+s_2-1,2p-1-s_1-s_2),
\end{aligned}
\end{equation}
and the summations over $k$ and $l$ are incremented in steps of $\Delta k = \Delta l = 2$.

The simplest example is the Ising model $\mathcal{M}(4,3)$. In our notation the correspondence between fields is as follows:
\begin{equation}
    \begin{aligned}
    \Psi_{1,1}\quad or\quad \Psi_{3,2} \quad &\Leftrightarrow \quad \mathbf{1}~,\\
    \Psi_{2,1}\quad or\quad\Psi_{2,2}\quad &\Leftrightarrow \quad \sigma~,\\
    \Psi_{1,2}\quad or\quad \Psi_{1,3}\quad &\Leftrightarrow \quad \epsilon~,
    \end{aligned}
\end{equation}
with fusion rules
\begin{equation}
    \begin{aligned}
    \sigma\times\sigma&=\mathbf{1}+\epsilon~,\\
        \sigma\times\epsilon&=\sigma~,\\
        \epsilon\times\epsilon&=\mathbf{1}~.
    \end{aligned}
\end{equation}
\section{Thimble representation of conformal blocks}
\label{sec:thimbles}
Now we want to compute the conformal blocks of the product of degenerate fields in Liouville theory:
\begin{equation}
    \begin{aligned}
    \langle \prod_a V_{-k_a/2b}(z_a)\rangle,
    \end{aligned}
\end{equation}
where $V_{-k_a/2b}=\Psi_{1,k_a+1}$. For general values of $b$ and $k_a$, they have the following fusion rules,
\begin{equation}
    \begin{aligned}
    V_{\frac{-k_1}{2b}}\times V_{\frac{-k_2}{2b}}=\sum_{l=|k_1-k_2|}^{k_1+k_2}V_{\frac{-l}{2b}}.
    \end{aligned}
\end{equation} This can be used to determine the dimension of the conformal block space.  

Our method to compute the conformal block is the free field realization also known as the Coulomb gas formalism\cite{Gaiotto:2011nm}.
Consider first a free field correlation function $\mathcal{F}$ with extra insertions of $V_{1/b}(w_i)$:
\begin{equation}
\label{eq:free}
    \begin{aligned}
    \mathcal{F}\left(\prod_iV_{1/b}(w_i)\prod_aV_{-k_a/2b}(z_a)\right)&=\prod_{i<j}(w_i-w_j)^{\frac{-2}{b^2}}\prod_{i,a}(w_i-z_a)^{\frac{k_a}{b^2}}\prod_{a<b}(z_a-z_b)^{\frac{-k_ak_b}{2b^2}}\\
    &=\exp\left(\frac{1}{b^2}\mathcal{W}(w,z)\right),
    \end{aligned}
\end{equation}
where $\mathcal{W}$ is known as the Yang-Yang superpotential due to its relation to 2d $\mathcal{N}=(2,2)$ Landau-Ginzburg models and integrable systems \cite{Jeong:2018qpc,Yang:1968rm}. Then the integrations 
\begin{equation}
\label{block}
    \langle\prod V_{-k_a/2b}(z_a)\rangle_\Gamma=\int_\Gamma \mathcal{F}\left(\prod_iV_{1/b}(w_i)\prod_aV_{-k_a/2b}(z_a)\right) \prod_idw_i,
\end{equation}
are the conformal blocks of the corresponding correlation functions. We may have more than one integration variable, denoted by $w_i$, and the integration cycle $\Gamma$ can be rather freely chosen as long as the integral converges. 

A particular choice of $\Gamma$ is given in terms of Lefschetz thimbles. Lefschetz thimbles are unions of the points on steepest descent paths flowing out from the critical points of $\mathcal{W}(w,z)$. More precisely, given a critical point $\sigma$ of $\mathcal{W}(w,z)$, the associated Lefschetz thimble is the union of all the endpoints $w(0)$ of paths $w(t),\; t\in (-\infty,0]$ which solve the equation:
\begin{equation} \label{eq:flow}
\begin{gathered} 
    \frac{d\bar w}{dt}=-\frac{\partial \mathcal{W}}{\partial w},\;\frac{d w}{dt}=-\frac{\partial\bar{\mathcal{W}}}{\partial\bar w},\\
     w(-\infty)=\sigma.
\end{gathered}
\end{equation}
We will use $\mathcal{J}$ to denote the thimble hereafter.
Note that, the position of the critical point $\sigma$ and the thimble $\mathcal{J}$ depend on the value of $z$. By employing such Lefschetz thimbles, the corresponding conformal blocks admit an interpretation as Landau-Ginzburg wavefunctions corresponding to a state $\alpha$ specified by the thimble \cite{VafaFQHE,Bergamin:2019dhg}:
\begin{equation}
    \psi_{\alpha} \equiv \int_{\Gamma_{\alpha}} \exp\left(\frac{1}{b^2}\mathcal{W}(w,z)\right) \prod_idw_i = \langle\prod V_{-k_a/2b}(z_a)\rangle_{\Gamma_{\alpha}}.
\end{equation}

In order to understand these thimbles and their properties better, let us separate the real and complex parts of the variable $w$ as $w=\mathfrak{Re}w+i \mathfrak{Im}w$. Then the flow equations \eqref{eq:flow} become
\begin{equation}
	\label{downward flow}
	\frac{d\mathfrak{Re}w}{dt}=-\mathfrak{Re}(\frac{\partial\mathcal{W}}{\partial w}),\quad \frac{d\mathfrak{Im}w}{dt}=\mathfrak{Im}(\frac{\partial\mathcal{W}}{\partial w}).
\end{equation}

If there is only one variable $w$ then for each $\sigma$ there are $2$ paths which satisfies $w(-\infty)=\sigma$ and the thimble is the joining of these $2$ paths.
From simple complex analysis, we know that $\mathfrak{Im}\mathcal{W}$ is conserved along the thimble, and is hence independent of the parameter $t$. Also, the integration over thimbles are guaranteed to converge because $\mathfrak{Re}\mathcal{W}\rightarrow -\infty$ when approaching the asymptotic regions.

Now we would like to prove a property of critical points and thimbles for later use. Given an integral of $w$
\begin{equation}
\label{eq:beforechange}
    \int \mathcal{W}(w,z)dw,
\end{equation}
we can change the integration variable to $y$:
\begin{equation}
    \eqref{eq:beforechange}=\int \mathcal{V}(y,z)dy,
\end{equation}
where 
\begin{equation}
    \mathcal{V}(y,z)=\mathcal{W}(w(y),z)\frac{dw}{dy}.
\end{equation}
Taking the derivative of $\mathcal{V}$ with respect to $y$ gives
\begin{equation}
\frac{\partial\mathcal{V}}{\partial y}=\frac{\partial \mathcal{W}}{\partial w}\cdot \left(\frac{dw}{dy}\right)^2+\mathcal{W}(w(y))\cdot\frac{d^2w}{dy^2}.
\end{equation}
If $w\rightarrow y(w)$ is a linear transformation, then for a critical point $\sigma$ of $\mathcal{W}$ we have 
\begin{equation}
(\frac{d^2 w}{dy^2}=0,\;\frac{\partial\mathcal{W}}{\partial w}|_\sigma=0 )\rightarrow\frac{\partial\mathcal{V}}{\partial y}|_{y(\sigma)}=0.
\end{equation}

This means, $y(\sigma)$ is a critical point of $\mathcal{V}$. Further, let $dw/dy=p$, then $\frac{\partial\mathcal{
V}}{\partial y}=p^2\frac{\partial \mathcal{W}}{\partial w}$. If $w(t)$ is a thimble parametrized by $t$, let $y(w(t))$ be the image of this thimble after the coordinate transformation. On this image, we have 
\begin{equation}
\begin{aligned}
        \frac{d\bar y}{dt}&=\frac{d\bar y}{d\bar w}\frac{d\bar w}{dt}\\
        &=-\frac{1}{p^*}\frac{\partial\mathcal{W}}{\partial w}\\
        &=-\frac{1}{p^*p^2}\frac{\partial\mathcal{V}}{\partial y}.
\end{aligned}
\end{equation}
And this implies
\begin{gather}
\frac{d\bar y}{d(t/p^2p^*)}=-\frac{\partial\mathcal{V}}{\partial y}.
\end{gather}
The complex conjugated equation can be derived similarly. So $y(w(t'))$ with $t'=t/(p^2p^*)$ is a steepest descent flow for $\mathcal{V}(y)$. We can conclude that under a linear coordinate transformation, the thimbles are preserved. 

In \eqref{eq:free}, $\mathcal{W}$ may be modified:
\begin{equation}
    \mathcal{W}(w,z)\rightarrow \mathcal{W}(w,z)+\Lambda(\sum_i w_i-\frac12\sum_a z_a),
\end{equation}
where $\Lambda$ is a constant independent of $w$ and $z$. This corresponds to inserting an irregular vertex operator at infinity. The added term is called the symmetry breaking term. The conformal blocks with irregular operators satisfy modified null vector identities and bear an irregular singularity at infinity themselves~\cite{Gaiotto:2012sf}. In this case we can simply regard the vacuum at infinity as the vector $\bra{0}$ which is then changed into an irregular vector $\bra{I}$. The fusion rule is not changed essentially as compared to the regular case without the irregular singularities, and we will say more on this in specific examples. 

Finally, we summarize the count of conformal blocks as deduced in \cite{Gaiotto:2011nm}. Consider a $d$ point degenerate field insertion. In the presence of the symmetry breaking term, there will be $\binom{d}{q}$ thimbles if the integration is $q$-fold ($q<d$). 
\subsection{The Stokes phenomena}\label{The Stokes phenomena}
We would like to examine the integration,
\begin{align}
    \int_{\mathcal{J}(z)} \exp\left(\frac{1}{b^2}\mathcal{W}(w,z)\right)dw,
\end{align}
in detail. Along the way, we will explain the so-called "Stokes phenomena" where for a more general and illustrative account we refer to \cite{inproceedings}. In the current exposition we will be following reference \cite{Witten:2010cx} which is closer to the relevant physical setup. 

Given a function $\mathcal{W}(w,z)$, we can solve \eqref{downward flow} to get the corresponding thimbles. This can be done numerically once we fix a particular complex number $z$, and the resulting critical points and thimbles can be drawn on the $w$ plane. When we slowly vary the value of $z$, the critical points always move accordingly in a smooth way on the $w$ plane. So it makes sense to denote them as $\sigma_i(z)$ with $i$ running from $1$ to the total number of critical points. For generic values of $z$, the corresponding thimble $\mathcal{J}_i(z)$ also deforms smoothly and we can get an analytical function in a small region using this variation. 

But it happens that, when $z$ crosses some "curves" on the complex plane, $\mathcal{J}_i(z)$ jumps drastically although $\sigma_i(z)$ only moves by a small amount. These "curves" are called Stokes lines.

To further explain what happens, we first assign a direction(represented by arrows in following figures) to each thimble suggesting the direction of line integrals. For generic values of $z$, thimbles always start or end at singular points of $\mathcal{W}$ or at infinity, and different thimbles may have the same starting or ending point. In cases we will explore in the following, crossing $z$ over a Stokes line will only change the end point of one of the thimbles, in a way such that the new thimble after all the jumps is the combined flow of $2$ old thimbles along arrows. 

The joining of the integral contours amounts to adding up two integrals. So we can easily see that when Stokes phenomena happen, the two sets of functions defined by thimble integration before and after the jumps differ by a linear transformation, namely the new set of homology classes of
thimbles are linear combinations of the old set. And this linear combination is represented by the Stokes matrices.

For brevity, below we will use "thimble" to refer to both the integration contour and the values of the integral(functions) over them. 

\begin{figure}
    \hspace*{-0.9cm}
\includegraphics[scale=.3,trim={10cm 18cm 12cm 6cm},clip]{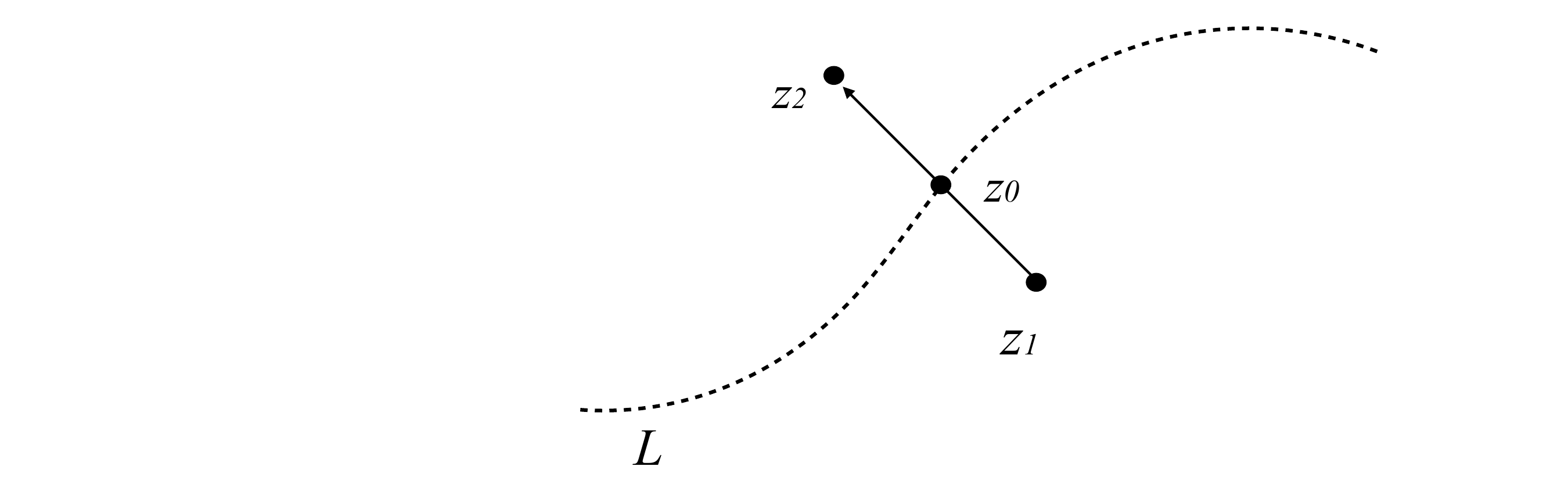}
    \caption{Stokes line, in $z$ plane}
    \label{crossingline}
\end{figure}
\begin{figure}
  \centering
  \begin{tabular}{ccc}                                                            
   \includegraphics[scale=0.2,trim={24cm 10cm 22cm 6cm},clip]{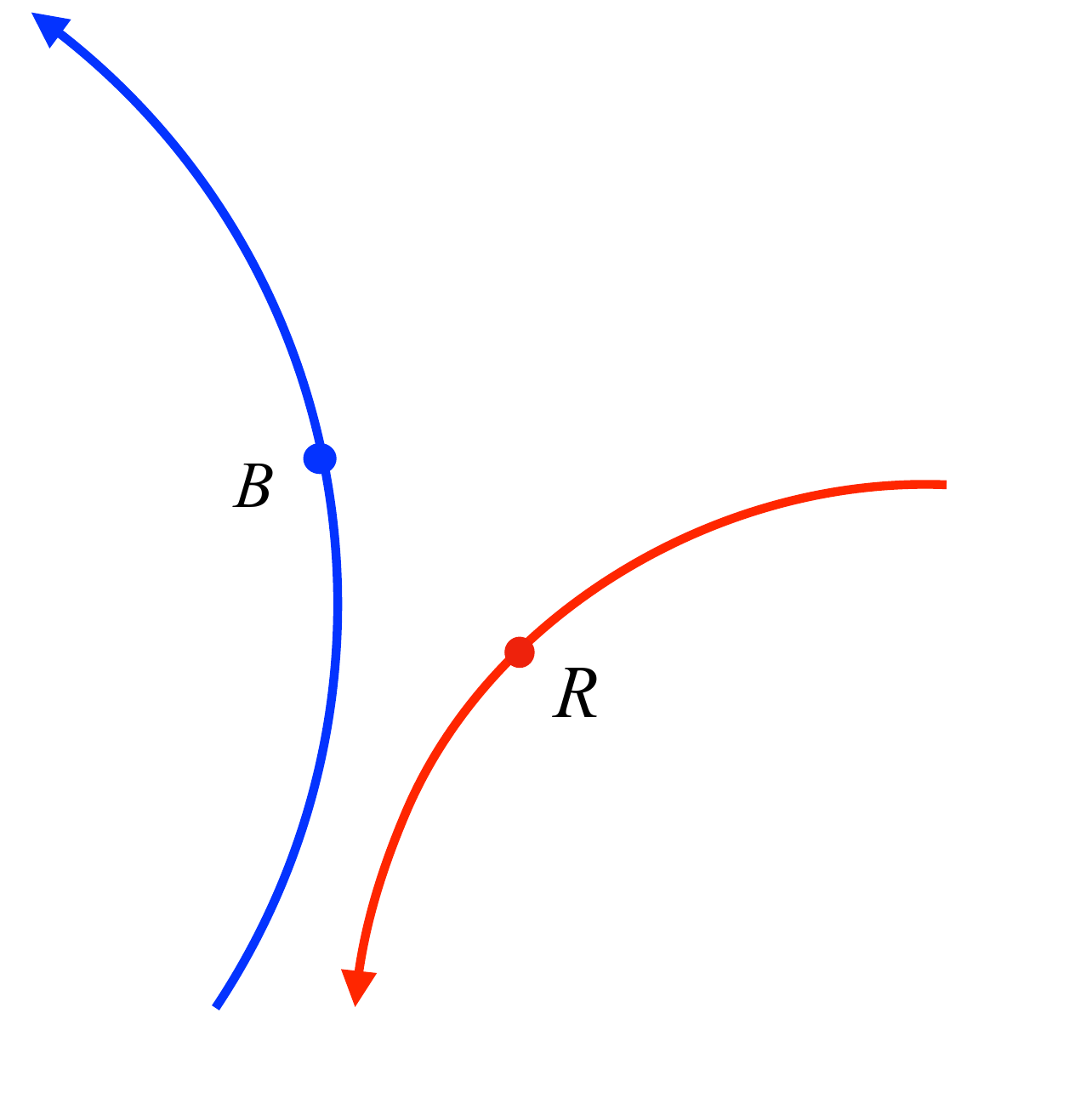}&        
    \includegraphics[scale=0.2,trim={25cm 10cm 20cm 8cm},clip]{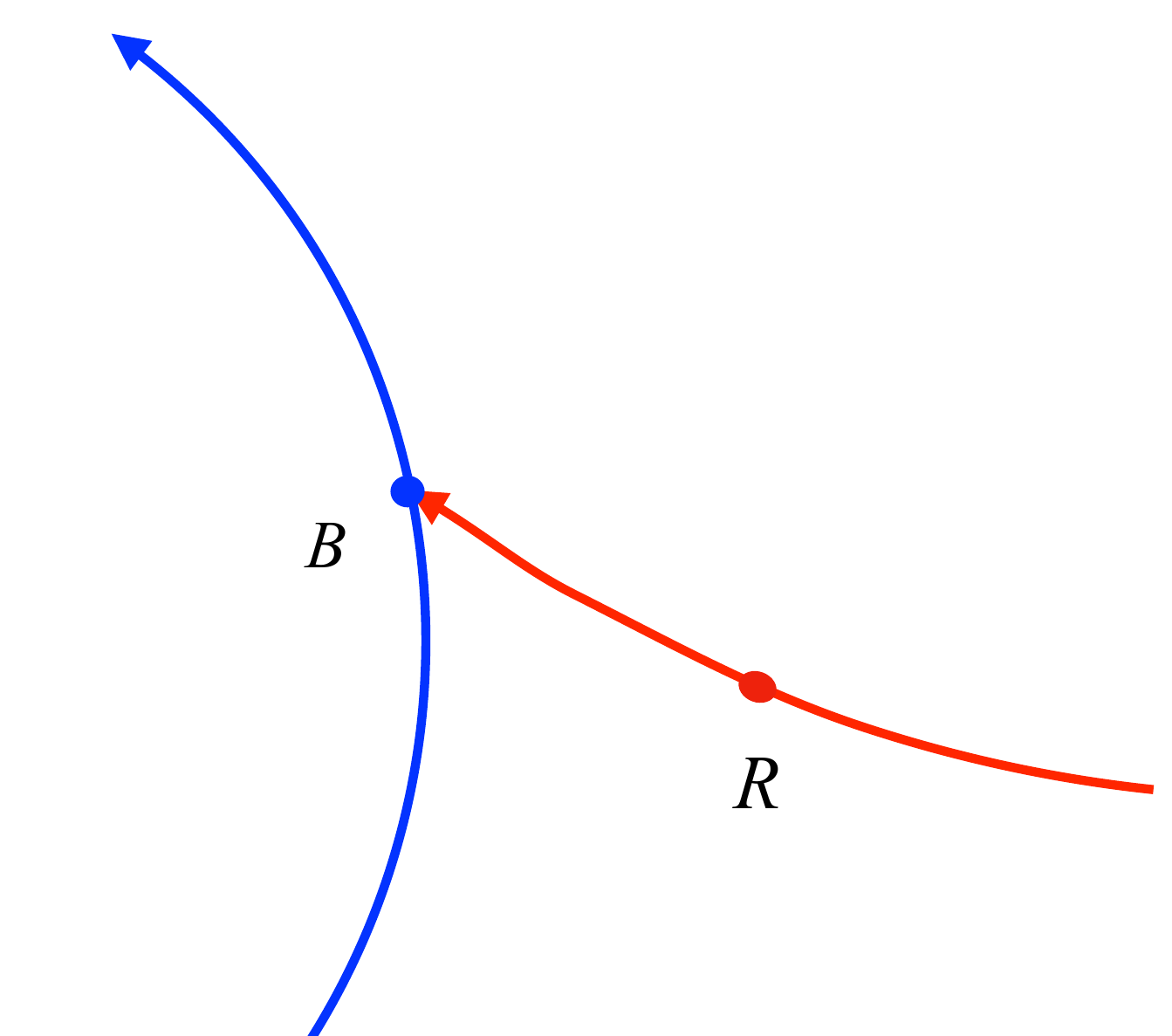}&             
    \includegraphics[scale=0.2,trim={25cm 10cm 20cm 6cm},clip]{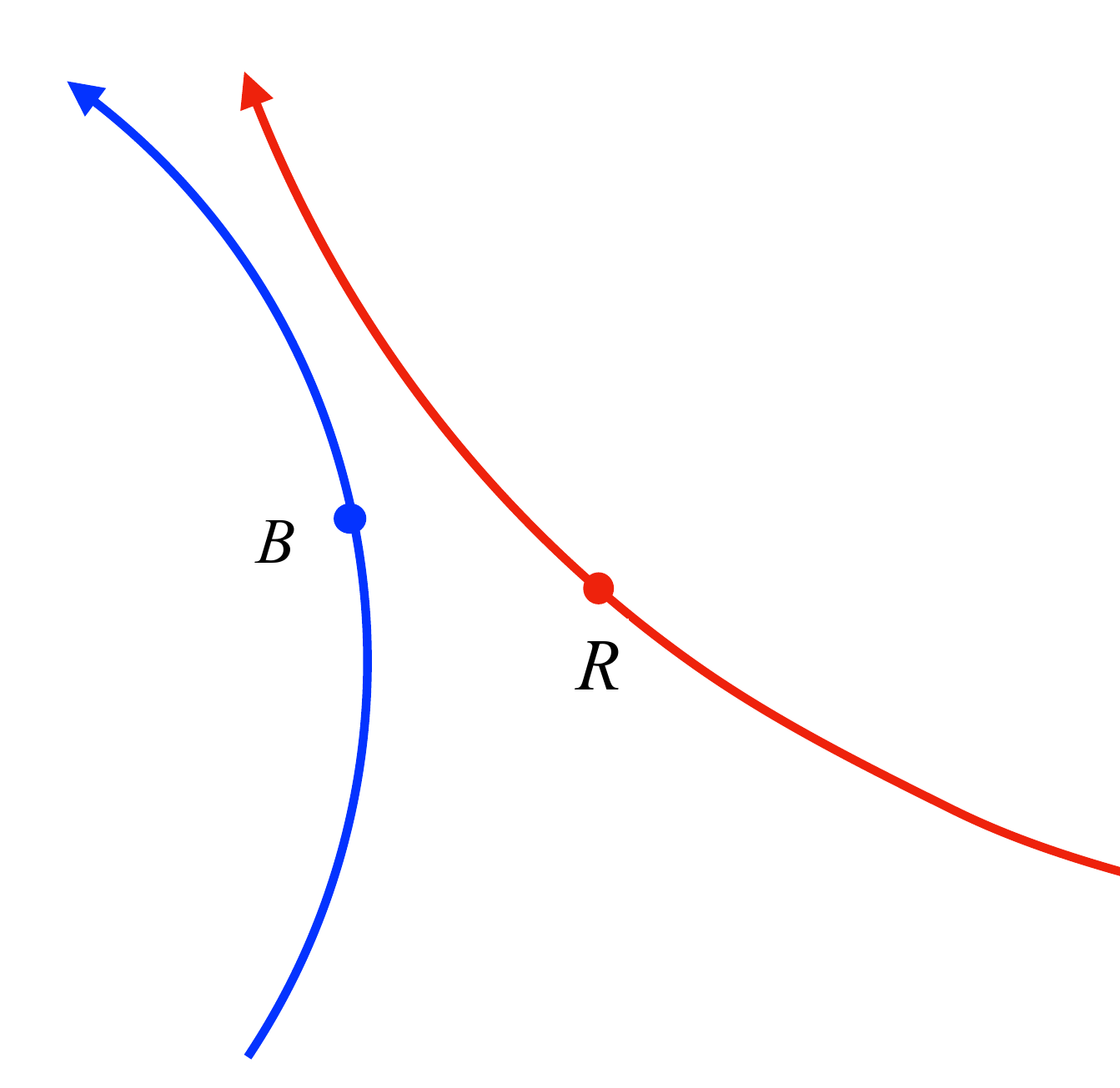}                               
  \end{tabular}
  \caption{From left to right are thimbles at $z=z_1,\;z_0,\;z_2$ respectively, in $w$ plane}
  \label{fig:Stokesjump}
\end{figure}

We now turn to \cref{crossingline,fig:Stokesjump}. There is a Stokes line $L$ lying on the $z$ plane. When $z=z_1$ which sits on downside of $L$, the thimbles are as in the leftmost one of \cref{fig:Stokesjump}.   We use $R$ and $B$ to denote 2 different critical points in the $w$ plane. And we label corresponding two thimbles as $\mathcal{J}_R$ and $\mathcal{J}_B$. when $z$ is varied to $z_0$ which sits exactly on $L$, $\mathcal{J}_R$ ends at $B$. This also means there exists a downward flow from $R$ to $B$. When $z$ is further varied to $z_2$ which sits on the upper side of $L$, the thimbles will become as in the rightmost one of \cref{fig:Stokesjump}. 

$\mathcal{J}_R$ after the jump($z=z_2$) is clearly the addition of $\mathcal{J}_B$ and $\mathcal{J}_R$ before the jump($z=z_1$). So the Stokes matrix $S$ associated with the Stokes line $L$ can be written in this thimble basis as follows,
 \begin{equation}
     S\begin{pmatrix}
      \mathcal{J}_R\\
      \mathcal{J}_B
     \end{pmatrix}
     =\begin{pmatrix}
     1&1\\
     0&1\end{pmatrix}
     \begin{pmatrix}
      \mathcal{J}_R\\
      \mathcal{J}_B
     \end{pmatrix}.
 \end{equation}

Sometimes, the unchanged thimble $\mathcal{J}_B$ has an opposite direction. In this case the Stokes matrix takes $-1$ in the upper-right corner.

There is a convenient way to determine the Stokes lines on the $z$ plane. When $z$ sits exactly on a Stokes line, like $z=z_0$ in \cref{crossingline}, this will be a downward flow from one of the critical point $i$ to the other $j$ (In \cref{fig:Stokesjump}, the flow is from $R$ to $B$). So the equation
\begin{equation}
\label{eq:Imaginarypart}
    \mathfrak
{Im}\mathcal{W}_i=\mathfrak
{Im}\mathcal{W}_j,
\end{equation} for two of the critical values $\mathcal{W}_i=\mathcal{W}(\sigma_i,z)$ and $\mathcal{W}_j=\mathcal{W}(\sigma_j,z)$, is satisfied, since $\mathcal{W}$ is conserved along the flow. The general strategy  is to list and solve all equations of the form \eqref{eq:Imaginarypart}, and draw the solution sets on the $z$-plane, which are indeed codimensional-$1$ lines. But one should keep in mind that for general $\mathcal{W}$ \eqref{eq:Imaginarypart} is only a necessary condition for the existence of downward flows between $\sigma_i$ and $\sigma_j$. There may be values of $z$ satisfying \eqref{eq:Imaginarypart} while not supporting any downward flow between critical points. We should pick up lines among solution sets the genuine Stokes lines. 

\begin{figure}
    \centering
    \includegraphics[scale=0.3,trim={15cm 10cm 15cm 10cm},clip]{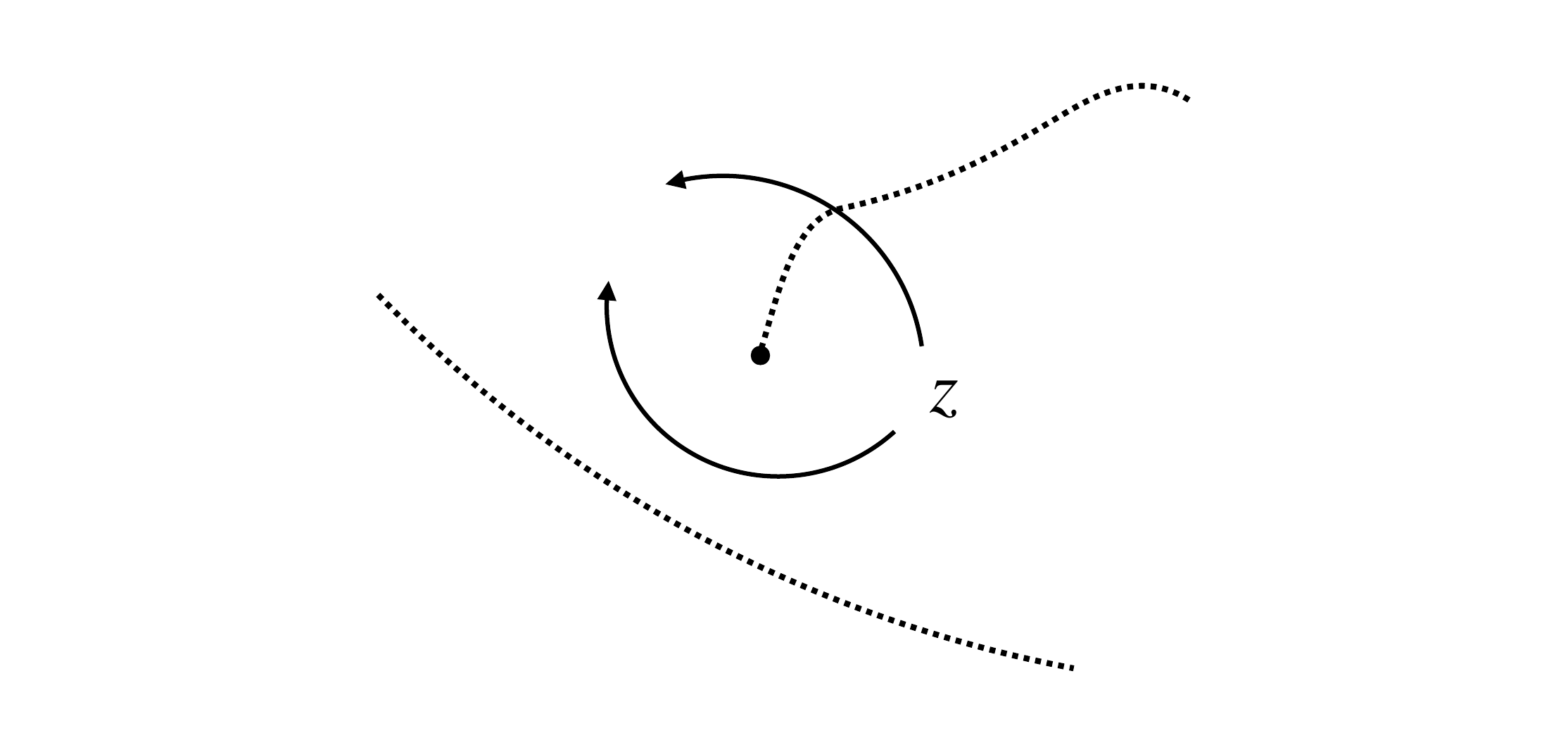}
    \caption{Varying $z$ in $2$ different paths denoted by solid lines causes an inconsistency.}
    \label{fig:Sort}
\end{figure}

Lines of the solution sets may intersect each other and form junctions. However, genuine Stokes lines cannot end on a point which is not junctions. Assume there is such a point, then varying $z$ through different paths would change thimbles in different ways, which is a contradiction. See \cref{fig:Sort}. 
So we can discard such lines from the solution sets. After this, the surviving lines can unambiguously divide the plane into sectors. The last step is to check whether there is indeed a jump of thimbles when crossing these lines. If no, we simply eliminate this line as well. If there is, we can go a step further to compute the corresponding Stokes matrix. 

We will refer to the sectors divided by Stokes lines as Stokes sectors. Each Stokes sector corresponds to a set of analytical functions that forms a basis for the space of conformal blocks.  

\subsection{Summary}
Now we have integral representations $\mathcal{I}_i$ of conformal blocks
\begin{equation}
\label{intconformalblock}
    \begin{aligned}
\mathcal{I}_i(z)=\int_{\mathcal{J}_i(z)} \exp\left(\frac{1}{b^2}\mathcal{W}(w,z)\right)dw,
    \end{aligned}
\end{equation}
where the $i$'s label the critical points of $\mathcal{W}$. As we have seen, this integral representation is only valid in certain Stokes sectors. Beyond the corresponding sectors, the conformal block can be described by the analytic continuation and we will keep using the notation $\mathcal{I}_i(z)$ in the extended region.

Suppose there are two adjacent Stokes sectors, labeled by $A$ and $B$. Associated to $A$ there is a conformal block basis $\{\mathcal{I}_i^A(z)\}$ and to $B$ there is $\{\mathcal{I}_i^B(z)\}$ associated. In other words, each set consists of different analytical functions in different Stokes sectors but associated to the ``same'' set of critical points. They are all dominated by contributions around that critical point, since they are steepest descent paths and the integrand is of exponential form. The leading contribution is proportional to the value of the integrand at the critical point $w_i$:
\begin{equation}
\label{eq:Saddle}
  \int_{\mathcal{J}_i(z)} \exp\left(\frac{1}{b^2}\mathcal{W}(w,z)\right)dw \sim C \exp\left(\frac{1}{b^2}\mathcal{W}(w_i,z)\right).
\end{equation}
where $C$ is a proportional coefficient.
This is valid regardless of which sector $z$ belongs to, so
\begin{equation}
\label{eq:pureformalmono}
    \begin{aligned}
\mathcal{I}_i^A(z)&\sim C \exp\left(\frac{1}{b^2}\mathcal{W}(w_i,z)\right),\quad z\in A,\\
\mathcal{I}_i^B(z)&\sim C \exp\left(\frac{1}{b^2}\mathcal{W}(w_i,z)\right),\quad z\in B,
    \end{aligned}
\end{equation}
with the same $C$.
This property of thimbles is useful in the computation of their monodromy. We assume $\sigma_i$ is isolated and non-degenerate when we use \eqref{eq:Saddle}. For $\mathcal{W}$ in the form of \eqref{eq:free}, the critical points are always isolated. There are cases when critical points coincide with each other and become degenerate. But this requires $z$ to take some special values and does not affect the computation of monodromies,  
\section{The monodromy of conformal blocks}
\label{sec:monodromy}
We are interested in the 
behavior of conformal blocks around their singularities. In particular, we want to compute the monodromy, which is an essential property of multivalued functions. We would like to emphasize here that from this section we will show something new, whereas materials in the previous sections are known in the literature.

If we analytically continue $z$ around a singular point, the $\mathcal{I}_i(z)$ become linear combinations of other $\mathcal{I}_j$'s. This change is encapsulated in the monodromy matrices $M$:
\begin{equation}
\label{eq:actualmono}
   \mathcal{I}_i(e^{2\pi i }z)=(M)_{ij} \mathcal{I}_j(z),
\end{equation}

For $\mathcal{I}(z)$ in integral form representation, there is a straightforward method to compute the analytical continuation. The essence is to deform the integration contour. We will elaborate on this method in the next section using concrete examples.  

\begin{figure}
    \centering
    \includegraphics[trim={0 3cm 10cm 5cm},clip,scale=.4]{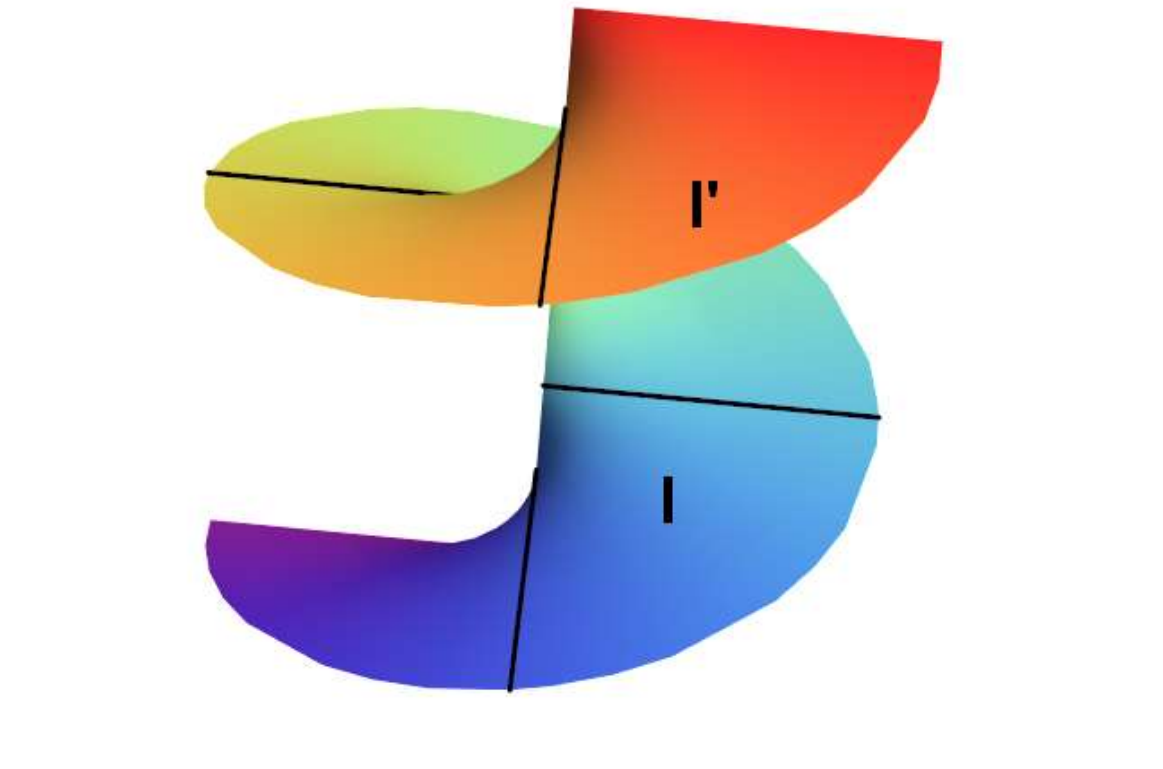}
    \caption{Branches $z=a$. The black solid lines represent Stokes lines}
    \label{fig:sheets}
\end{figure}

Apart from the contour deformation method, there is another way to represent the monodromy once we know the Stokes matrices. Recalling \cref{eq:Saddle}, one can further expand $\exp\left(\frac{1}{b^2}\mathcal{W}(w_i,z)\right)$ around the singular point $a$ to get a series $P_i(z-a)$ with leading term $(z-a)^{\rho_i}$. The singularities always have Stokes lines passing through them and this expansion is valid in all Stokes sectors surrounding $a$. When $\rho_i$ is not an integer, it happens that then the function is not single valued around $z=a$, and we must bear in mind a multi-sheet Riemann surface with Stokes lines on each sheet. The patterns of the Stokes lines on different branches are isomorphic, see \cref{fig:sheets}.
The sector $I'$ is exactly the lift of $I$, and thus $z-a$ in $I'$ differs from that in $I$ by multiplication by a factor $e^{2\pi i}$. 
From the argument above we have a relation between the conformal blocks $\{\mathcal{L}_i\}$ associated to $I$ and those associated to $I'$, denoted by $\{\mathcal{L}'_i\}$:
\begin{equation}
\label{eq:formalmono}
    \mathcal{I}'_i(e^{2\pi i}(z-a))=e^{2\rho_i\pi i}\mathcal{I}_i(z-a).
\end{equation}
The phase $e^{2\rho_i\pi i}$ is called the formal monodromy.
This is very much like \cref{eq:actualmono} except that the functions are from different bases. To proceed, we first rewrite \cref{eq:formalmono} in matrix form:
\begin{equation}
\label{eq:matrixformal}
    \begin{pmatrix}
    \mathcal{I}_1'(e^{2\pi i}z)\\
    \mathcal{I}_2'(e^{2\pi i}z)\\
    \vdots
    \end{pmatrix}=\hat{M}\begin{pmatrix}
    \mathcal{I}_1(z)\\
    \mathcal{I}_2(z)\\
    \vdots
    \end{pmatrix},
\end{equation}
where $\hat{M}$ is the diagonal formal monodromy matrix. Now we are ready to use Stokes matrices to transfer the left hand side of \eqref{eq:formalmono} back to the $\mathcal{I}_i$ basis. Let us denote the Stokes matrices we encounter, when performing analytic continuation, by $S^1$, $S^2$, ... , $S^p$, see \cref{fig:sheets}. Plugging these successive basis transformations into \cref{eq:matrixformal}, we get
\begin{equation}
 S^P...S^{2}S^{1}\begin{pmatrix}
    \mathcal{I}_1(e^{2\pi i}z)\\
    \mathcal{I}_2(e^{2\pi i}z)\\
    ...
    \end{pmatrix}=\hat{M}\begin{pmatrix}
    \mathcal{I}_1(z)\\
    \mathcal{I}_2(z)\\
    ...
    \end{pmatrix}.
\end{equation}
Or equivalently,
\begin{equation}
    \begin{pmatrix}
    \mathcal{I}_1(e^{2\pi i}z)\\
    \mathcal{I}_2(e^{2\pi i}z)\\
    ...
    \end{pmatrix}=(S^P...S^2S^1)^{-1}\hat{M}\begin{pmatrix}
    \mathcal{I}_1(z)\\
    \mathcal{I}_2(z)\\
    ...
    \end{pmatrix}.
\end{equation}
Thus the actual monodromy matrix is
\begin{equation}
\label{eq:master}
    M=(S^P...S^2S^1)^{-1}\hat{M}.
\end{equation}

\subsection{Example: Modified Bessel functions}
\begin{figure}
\centering
\subfloat[a][The regular conformal block]
{
\begin{tikzpicture}
\draw[black, thick] (-3,0)node[anchor=east]{$V_{-\frac{1}{2b}}(z_1)$} -- (3,0)node[anchor=west]{$V_{-\frac{1}{2b}}(\infty)$};
\draw[black, thick] (-1,0.7)node[anchor=south]{$V_{-\frac{1}{2b}}(z_2)$} -- (-1,0);
\draw[black, thick] (1,0.7)node[anchor=south]{$V_{-\frac{1}{2b}}(z_3)$} -- (1,0);
\node at (0,0.3) {$\Psi$};
\end{tikzpicture}
}
\\
\subfloat[b][The conformal block obtained by colliding $V(z_2)$ and $V(\infty)$]
{
\begin{tikzpicture}
	\draw[black, thick] (-3,0)node[anchor=east]{$V_{-\frac{1}{2b}}(z_1)$} -- (1,0);
	\draw[black, thick] (-1,0.7)node[anchor=south]{$V_{-\frac{1}{2b}}(z_2)$} -- (-1,0);
	\node at (0,0.3) {$\Psi$};
	\node at (2,0.3){$\alpha$};
	\filldraw [black] (1,0) circle (1.5pt);
 \path [draw,snake it]
    (1,0)--(3,0);
\end{tikzpicture}
}
\caption{Fusion of 2 point function with an irregular singularity of rank $\frac{1}{2}$}
\label{fig:fusionchannel}
\end{figure}
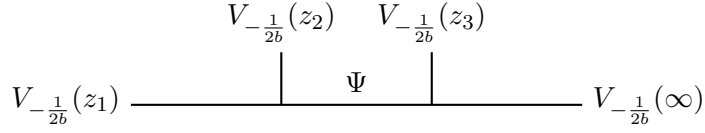
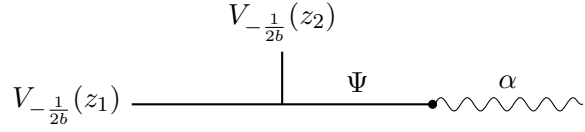

We now consider a simple case of \eqref{block} and \eqref{eq:free}: the $2$ point correlators with both $k_a=1$ and symmetry breaking. The relevant fusion rule reads
\begin{equation}
\label{eq:fusionrule}
    V_{-\frac{1}{2b}}\times V_{-\frac{1}{2b}}=V_{0}+V_{-\frac{1}{b}}.
\end{equation}
The irregular vector at infinity is created by colliding 2 degenerate fields, see \cref{fig:fusionchannel} where we are following the conventions of reference \cite{Bonelli:2022ten}.
In our case the singularity is of rank $\frac{1}{2}$, denoted by the wavy line. We have two choices of $\Psi$ as shown in \cref{eq:fusionrule}.
This means that our conformal block space is two-dimensional.
The relevant integral is 
\begin{equation}
\label{eq:intconformal}
\mathcal{I}=\int_{\mathcal{{J}}} (w-z_1)^{1/b^2}(w-z_2)^{1/b^2}(z_1-z_2)^{-1/2b^2}e^{\frac{\Lambda}{b^2}(\frac{z_1}{2}+\frac{z_2}{2}-w)}dw.
\end{equation}
After a coordinate transformation $W=\Lambda(w-z_2)$ and $Z=\Lambda(z_1-z_2)$, we get 
\begin{equation}
\label{integralrep}
	\mathcal{I}=
	(\frac{1}{\Lambda})^{1+3/2b^2}(\frac{1}{Z})^{1/2b^2}\int (W-Z)^{\frac{1}{b^2}}W^{\frac{1}{b^2}}(e^{\frac{Z}{2}-W})^{\frac{1}{b^2}}dW.
\end{equation}
The integration part can be rewritten in the following way:
\begin{align}
	\label{eq:integral}
	\begin{split}
\int(W-Z)^{1/b^2}W^{1/b^2}(e^{Z/2-W})^{1/b^2}dW&=\int[(W-Z/2)^2-\frac{Z^2}{4}]^{1/b^2}(e^{Z/2-W})^{1/b^2}dW\\	&=\int(Y^2-\frac{Z^2}{4})^{1/b^2}e^{-\frac{Y}{b^2}}dY\\
&=(\frac{Z}{2})^{\frac{2}{b^2}+1}\int_{\mathcal{J}'}(t^2-1)^{1/b^2}e^{-\frac{Z}{2b^2}t}dt
	\end{split},
\end{align}
where we have performed coordinate transformations $Y=W-\frac{Z}{2}$ and $t=\frac{2}{Z}Y$. We eventually change the integral variable from $w$ to $t$, which is a linear transformation. So the integral contour $\mathcal{J}'$ in \eqref{eq:integral} is still a thimble.

One of the integral representations of the modified Bessel function $K_\nu(x)$ is \cite{abramowitz+stegun}
\begin{equation}
\label{eq:besselint}
    	K_\nu(x)=\frac{\sqrt{\pi}(\frac12x)^\nu}{\Gamma(\nu+\frac12)}\int_{1}^{\infty}(t^2-1)^{\nu-\frac12}e^{-2xt}dt,\qquad |\arg x|<\frac{\pi}{2}.
\end{equation}
We note that one of the thimbles $\mathcal{J}'$ in \eqref{eq:integral} is $(1,\infty)$ which can be seen from setting $x\in\mathbf{R}^+$, and numerically solving \eqref{downward flow}. 
Comparing now \eqref{integralrep} with \eqref{eq:besselint}, we see that \eqref{integralrep} can be written as 
\begin{equation}
   \label{eq:relation} \mathcal{I}_{1}=Cx^{\frac{1}{2b^2}+\frac{1}{2}}K_{\frac{1}{b^2}+\frac{1}{2}}(x),
\end{equation}
with $x=\frac{Z}{4b^2}$ and constant $C=\frac{\sqrt{2}\Gamma(1+\frac{1}{b^2})}{\sqrt{\pi}c^{1+\frac{3}{2b^2}}}b^{1+\frac{3}{b^2}}$.

\begin{figure}
    \centering

    \includegraphics[scale=0.5]{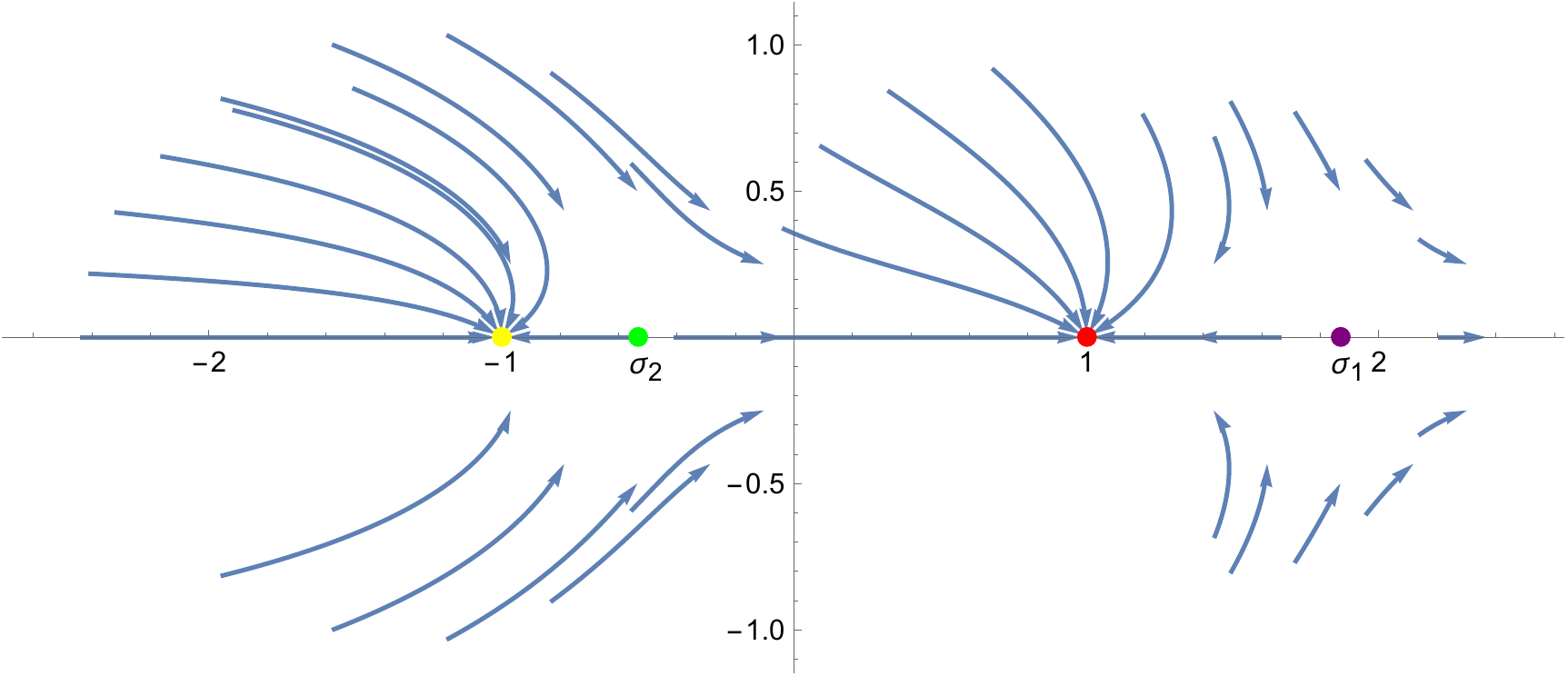}
    \caption{Downward flow by the integrand in \cref{eq:besselint} when $x=3$. \\ Note this is in $t$-plane.}
    \label{fig:modbesselthim}
\end{figure}

Let the integrand in \eqref{eq:besselint} be 
$f(t)=(t^2-1)^{\nu-\frac12}e^{-2xt}$.
Here and below we will assume $b$ to take an imaginary value, so that $1/b^2$ is negative. Then let $x=3$, and draw the downward flows determined by $f(t)$ in the $t$-plane. The result is shown in \cref{fig:modbesselthim}. The curves(lines) with arrows denote downward flows. $+1$ and $-1$ are two singularities of $f(t)$, while $\sigma_1$ and $\sigma_2$ are two critical points. The first thimble $\mathcal{J}_1$ is simply the line $(1,+\infty)$, passing through $\sigma_1$.
Also we can see another thimble, which flows from $+1$ to $-1$, passing through $\sigma_2$. This is related to another modified Bessel function:
\begin{equation}
    	I_\nu(x)=\frac{(\frac12x)^\nu}{\sqrt{\pi}\Gamma(\nu+\frac12)}\int_{-1}^{1}(1-t^2)^{\nu-\frac12}e^{-2xt}dt,\qquad \mathfrak{Re}\nu>-\frac{1}{2}.
\end{equation}
In other words
\begin{equation}
    \mathcal{I}_{2}= (-1)^{\frac{1}{b^2}}\pi C x^{\frac{1}{2b^2}+\frac{1}{2}}I_{\frac{1}{b^2}+\frac{1}{2}}(x).
\end{equation}
with $x$ and $C$ the  same as $\eqref{eq:relation}$.

We managed to identify our conformal blocks \eqref{eq:intconformal} with Bessel-like functions. In fact, the monodromy itself can be directly read off from the analytic continuation properties of these special functions. The formulae we need are\cite{abramowitz+stegun}
\begin{equation}
\begin{aligned}
	K_\nu(xe^{m\pi i})&=e^{-m\nu\pi i}K_\nu(x)-\pi i\sin(m\nu \pi)\csc(\nu\pi)I_\nu(x), \\   
    I_\nu(xe^{m\pi i})&=e^{m\nu \pi i}I_\nu(x).
\end{aligned}
\end{equation}
From this we obtain the monodromy around $x=0$:
\begin{equation}
   M_0\begin{pmatrix}
    \mathcal{I}_{1}\\
    \mathcal{I}_{2}
    \end{pmatrix}
    = \begin{pmatrix}
    e^{\frac{-\pi i}{b^2}}&e^{\frac{-\pi i}{b^2}}-e^{\frac{3\pi i}{b^2}}\\
    0&e^{\frac{3\pi i}{b^2}}
    \end{pmatrix}
    \begin{pmatrix}
    \mathcal{I}_{1}\\
    \mathcal{I}_{2}
    \end{pmatrix}.
\end{equation}
If we let $q=\exp(-\frac{2\pi i}{b^2})$, then \begin{equation}
    M_0=\begin{pmatrix}
        q^{1/2}&q^{1/2}-q^{-3/2}\\
        0&q^{-3/2}    
        \end{pmatrix},
\end{equation}
with eigenvalues $q^{-3/2}$ and $q^{1/2}$. This serves as a benchmark for our following method of computing monodromies.
\subsection{Stokes phenomena of modified Bessel functions}
We have proved that, under linear coordinate transformations, the critical points and thimbles are preserved. So we can choose instead \eqref{integralrep} as the integral representation. This is equivalent to taking the superpotential $\mathcal{W}$ to be(here we use $w$ and $z$ back instead of $W$ and $Z$ for simplicity) 
\begin{align}
\mathcal{W}=-\frac12\log z+\log(w-z)+\log w+\frac{z}2-w.
\end{align}
This superpotential has two critical points,
\begin{equation}
\label{eq:cri}
\begin{aligned}
\sigma_1=\frac12(2+z+\sqrt{4+z^2}),\quad 
\sigma_2=\frac12(2+z-\sqrt{4+z^2}).
\end{aligned}
\end{equation}
The critical values of the integrand are 
\begin{equation}
\begin{aligned}
    \exp(\mathcal{W}_1)&=\exp(-1-\frac{\sqrt{4+z^2}}{2})\frac{2+\sqrt{4+z^2}}{\sqrt{z}}
	\sim z^{-1/2},\\
\exp(\mathcal{W}_2)&=\exp(-1+\frac{\sqrt{4+z^2}}{2})\frac{2-\sqrt{4+z^2}}{\sqrt{z}}
	\sim z^{3/2},
\end{aligned}
\end{equation}
where $\sim$ denotes the leading exponent of the series expansion around $z=0$. This is the saddle point approximation of the original integration $\mathcal{I}_{i}$, and
the corresponding formal monodromy matrix is
\begin{equation}
    \hat{M}_0=\begin{pmatrix}
    q^{1/2}&0\\
    0&q^{-3/2}
    \end{pmatrix}.
\end{equation}

\begin{figure}
    \centering
\includegraphics[scale=0.5]{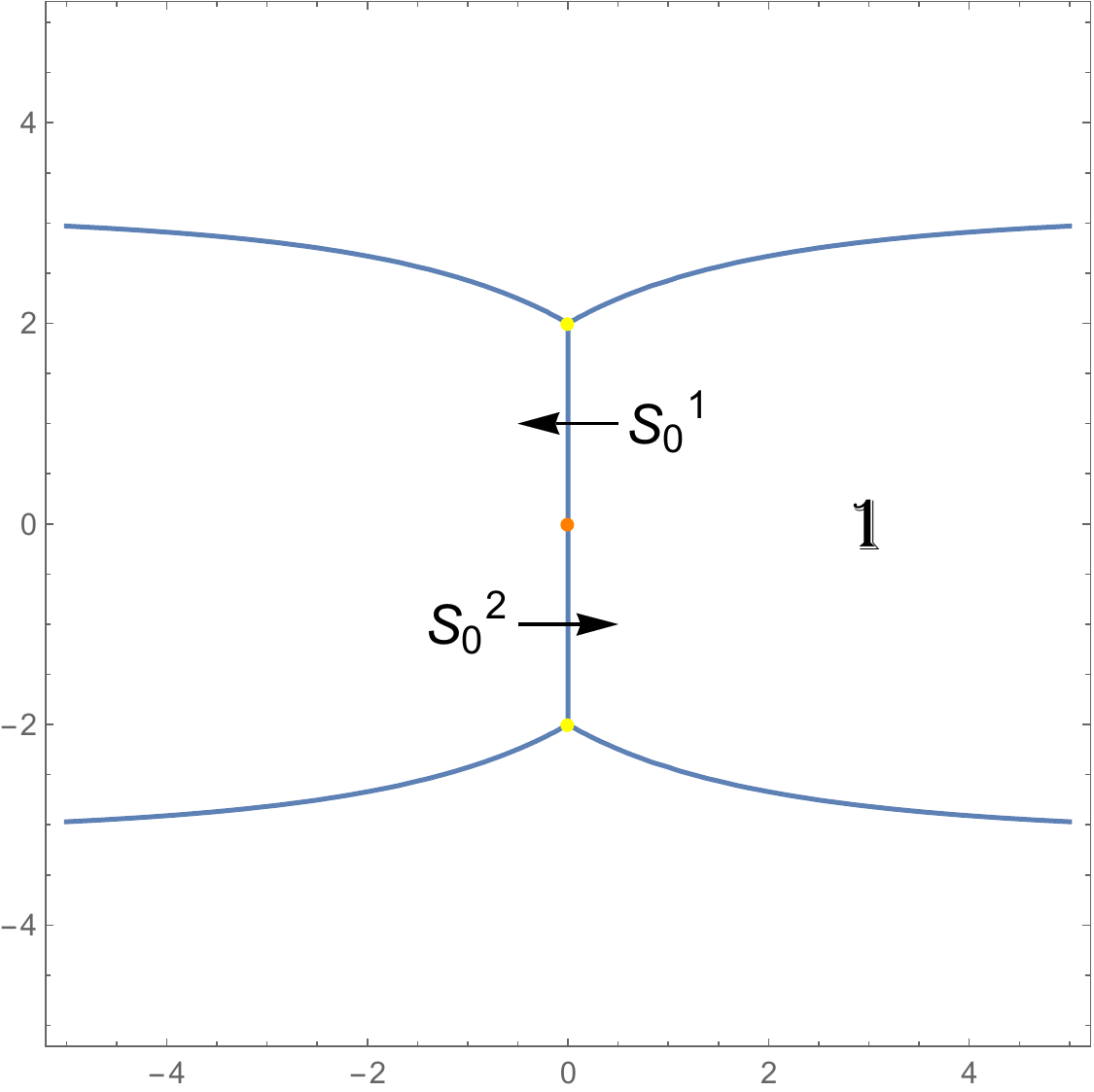}
    \caption{The Stokes lines corresponding to the integrand.\\
    This is in the $z$-plane}
    \label{fig:besselstokes}
\end{figure}

The next thing to do is to divide the $z$-plane into Stokes sectors using \eqref{eq:Imaginarypart}, see \cref{fig:besselstokes}.
Note that the modified Bessel functions are multivalued around the branch point $z=0$, so this figure is only the projection onto one of its branches. Also note that, there are two points $z=2i,-2i$ where $3$ Stokes lines meet. At these points $\sigma_1=\sigma_2$ and the saddle point approximation fails. But as long as we work in the right sector, this will do no harm to the argument. 

\begin{figure}
    \centering
\includegraphics[scale=0.3,trim={20cm 16cm 20cm 11cm},clip]{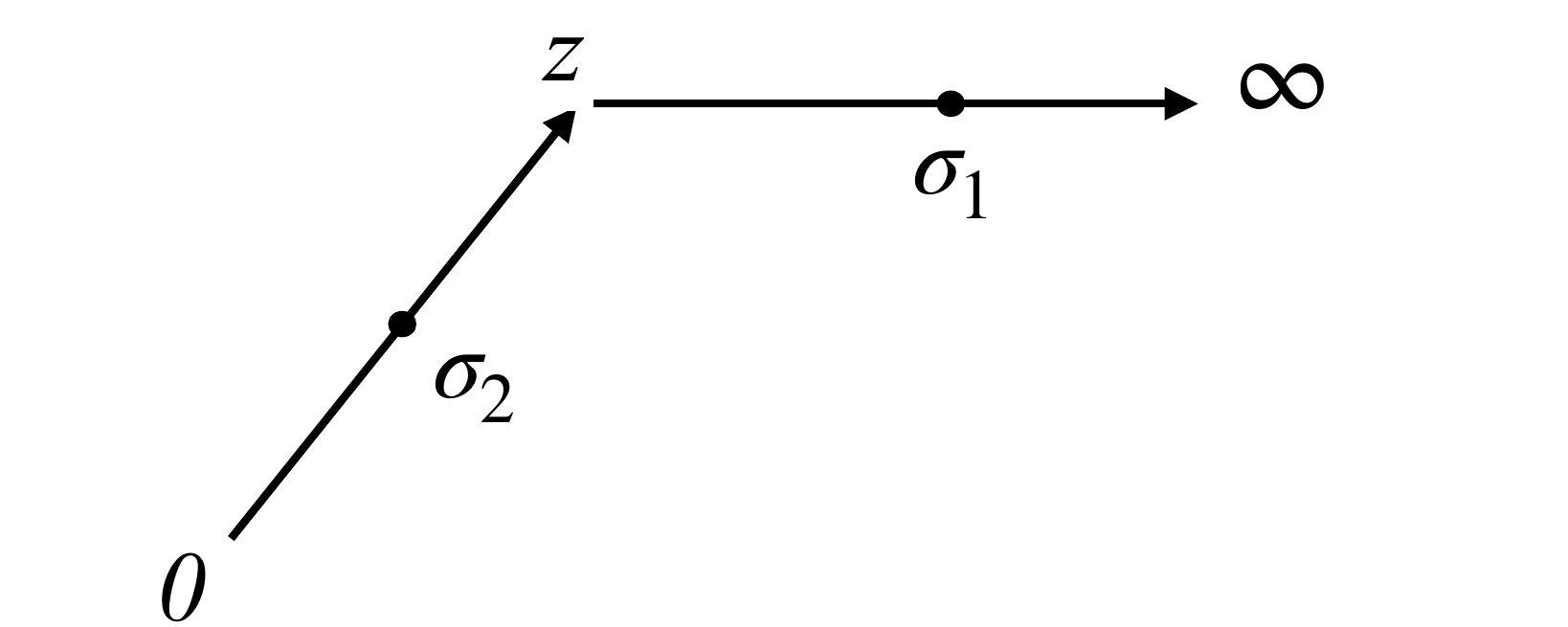}
    \caption{Thimbles in region $1$}
    \label{fig:besselthimble}
\end{figure}

Below we will work in the right-middle sector $1$ in \cref{fig:besselstokes}. This is equivalent to picking a basis of analytic functions for representing monodromy matrices. With $z$ in this region, the thimbles are the flows $\mathcal{J}_1=z\rightarrow \sigma_1\rightarrow \infty$ and $\mathcal{J}_2=0\rightarrow \sigma_2\rightarrow z$, respectively\footnote{We have chosen the labels such that they are consistent with \cref{eq:cri}.}, see \cref{fig:besselthimble}. The actual thimbles may not be straight lines, but here we represent them using straight lines for simplicity.

\begin{figure}
\centering
\begin{minipage}{.5\textwidth}
  \centering
  \includegraphics[scale=0.25,trim={22cm 9cm 25cm 8cm},clip]{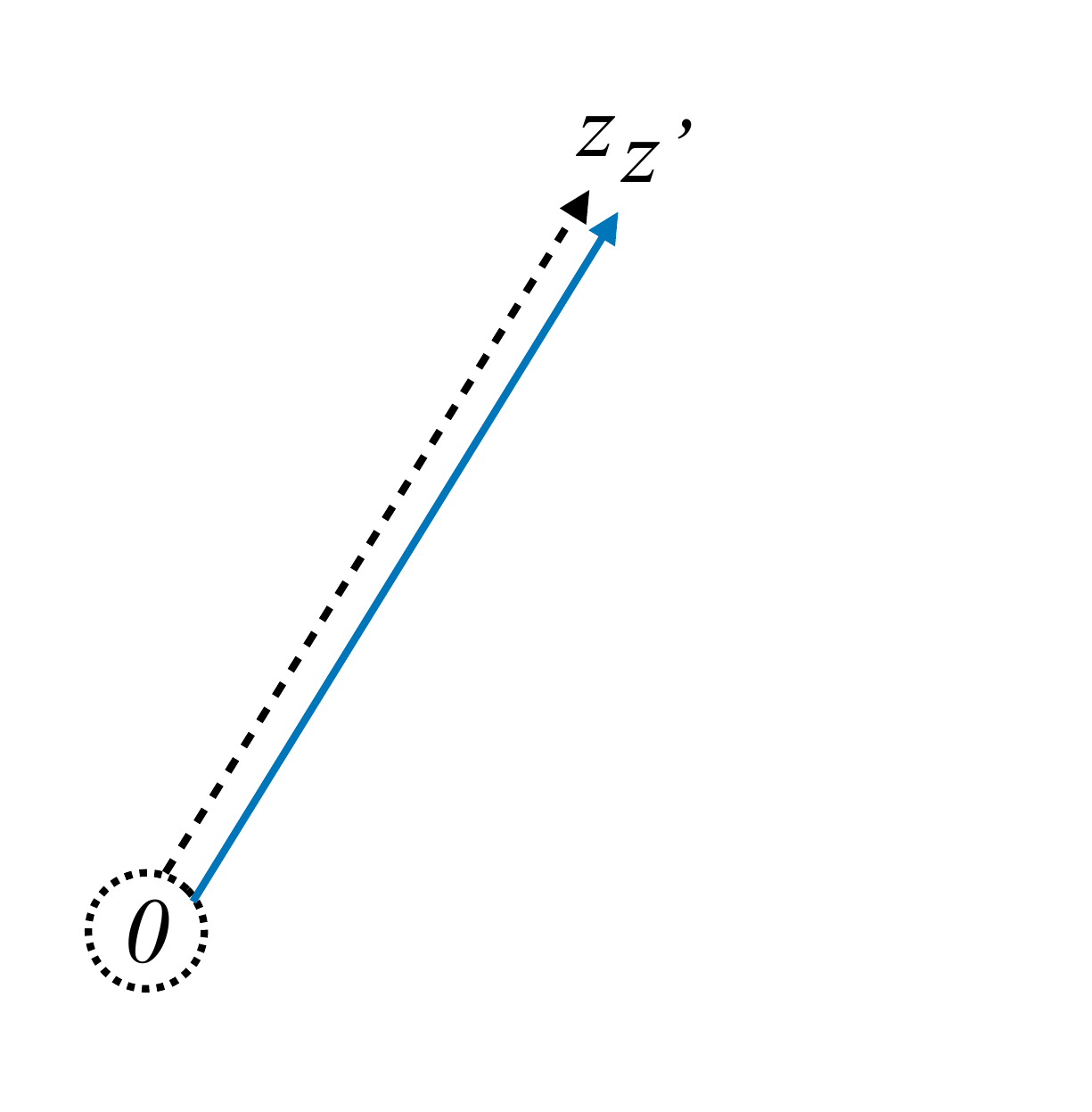}
  \captionof{figure}{$\mathcal{J}'_2$(
  The solid line) and \\$\mathcal{J}_2$(The dashed line)}
  \label{fig:realbesselthimble}
\end{minipage}%
\begin{minipage}{.5\textwidth}
  \centering
  \includegraphics[scale=0.25,trim={25cm 9cm 18cm 8cm},clip]{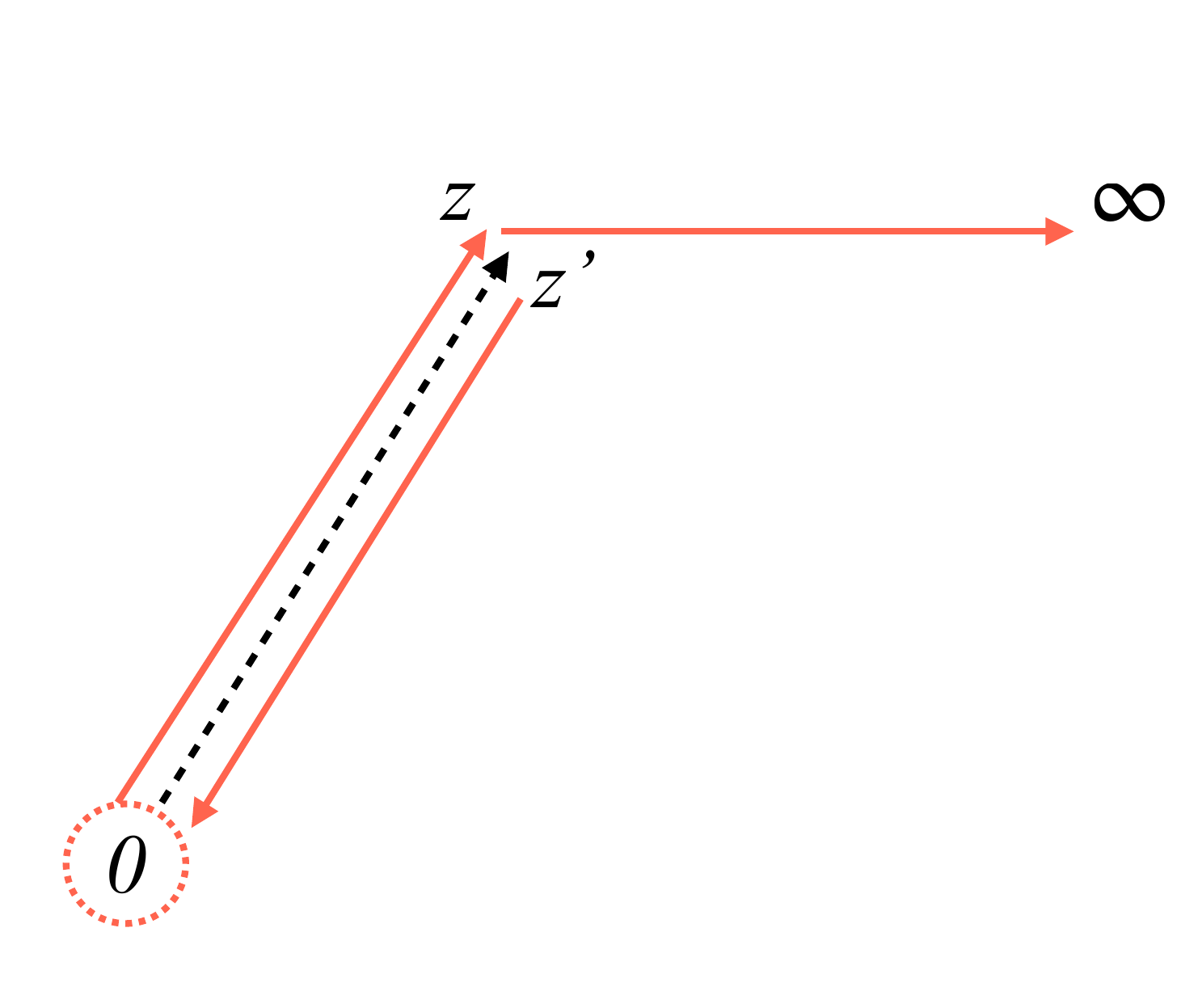}
  \captionof{figure}{$\mathcal{J}'_1$(The solid line) and\\ $\mathcal{J}_2'$(The dashed line)}
  \label{fig:realbesselthimbletwo}
\end{minipage}
\end{figure}

Now we would like to do the analytic continuation around $z=0$. Recall that, the relevant integrand is
\begin{equation}
\exp[\mathcal{W}/b^2]
=z^{-1/2b^2}w^{1/b^2}(w-z)^{1/b^2}e^{z/2b^2-w/b^2}.
\end{equation}
After the continuation, $z^{-1/2b^2}$ would pick up a phase: $z^{-1/2b^2}\rightarrow (z')^{-1/2b^2}=q^{1/2}z^{-1/2b^2}$. The contour $\mathcal{J}_2$ rotate counter-clockwisely around $0$ as a whole by $2\pi$ angle, ending on a different branch due to the multivaluedness of $\mathcal{W}$ around $0$. So on $\mathcal{J}_2$ the value of $w^{1/b^2}$ and $(w-z)^{1/b^2}$ also changes by $w^{1/b^2}\rightarrow q^{-1}w^{1/b^2}$ and 
$(w-z)^{1/b^2}\rightarrow q^{-1}(w-z)^{1/b^2}$. Now the total effect of the monodromy on $\mathcal{J}_2$ is 
\begin{equation}    
\mathcal{J}_2\rightarrow \mathcal{J}'_2=q^{-3/2} \mathcal{J}_2,
\end{equation}
see \cref{fig:realbesselthimble}.

$\mathcal{J}_1$ deforms in a more complicated way. The end point $\infty$ is fixed, while the starting point $z$ moves onto a different branch. From \cref{fig:realbesselthimbletwo} we can see that the contour is stretched into three pieces $-\mathcal
{J}_2'$, $\mathcal{J}_2$, and $\mathcal{J}_1$. So 
\begin{equation}
    \mathcal{J}_1'=(q^{1/2}-q^{-3/2})\mathcal{J}_2+q^{1/2}\mathcal{J}_1.
\end{equation}
The above results can be recapped in the following monodromy matrix: 
\begin{equation}
\label{eq:good}
M_0=\begin{pmatrix}
        q^{1/2}&q^{1/2}-q^{-3/2}\\
        0&q^{-3/2}
    \end{pmatrix}.
\end{equation}

We may also show the validity of \cref{eq:master}, which provides another perspective on the monodromy. A small circle around $z=0$ would cross two Stokes lines, as in \cref{fig:besselstokes}, and we label the corresponding Stokes matrices as $S_0$ and $S_0^1$, respectively. To determine these matrices, one may apply the method in \cref{The Stokes phenomena}. But there we ignored the multivaluedness of $\mathcal{W}$. Reconsidering it, we obtain the correct Stokes matrices:  
\begin{equation}
\begin{gathered}	S_0^1=\begin{pmatrix}
		1&1\\
		0&1
	\end{pmatrix},\;
	S_0^2=\begin{pmatrix}
		1&-q^2\\
		0&1
	\end{pmatrix}\\
\end{gathered}.
\end{equation}
In order for the basis after two jumps to have the correct leading behaviour \cref{eq:pureformalmono}, the integration contour of that basis must have contributions from two different sheets. This explains the appearance of $q^2$ in $S_0^2$. Now we have
\begin{equation}
    \begin{aligned}
M_0=(S_0^2S_0^1)^{-1}\hat{M}_0=\begin{pmatrix}
    1&-1\\
    0&1
    \end{pmatrix}
    \begin{pmatrix}
       1&q^2\\
    0&1
    \end{pmatrix}
    \begin{pmatrix}
     q^{1/2}&0\\
    0&q^{-3/2}
    \end{pmatrix}
    =\begin{pmatrix}
        q^{1/2}&q^{1/2}-q^{-3/2}\\
        0&q^{-3/2}    
        \end{pmatrix}
    \end{aligned},
\end{equation}
 consistent with \cref{eq:good}.

\subsection{Three point function with symmetry breaking term}
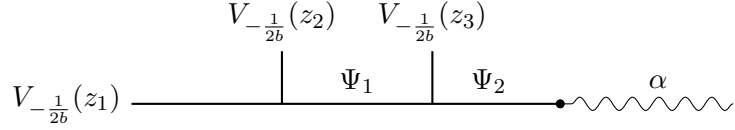
\begin{figure}
	\centering
 	\begin{tikzpicture}
		\draw[black, thick] (-4,0)node[anchor=east]{$V_{-\frac{1}{2b}}(z_1)$} -- (1.7,0);
		\draw[black, thick] (-2,0.7)node[anchor=south]{$V_{-\frac{1}{2b}}(z_2)$} -- (-2,0);
		\draw[black,thick](0,0.7)node[anchor=south]{$V_{-\frac{1}{2b}}(z_3)$} -- (0,0);
		
		\node at (-1,0.3) {$\Psi_1$};
		\node at (0.75,0.3) {$\Psi_2$};
		\node at (3,0.3){$\alpha$};
		\filldraw [black] (1.7,0) circle (1.5pt);
  \path [draw,snake it](4,0)--(1.7,0);
	\end{tikzpicture}
 \caption{Fusion of the 3 point function with an irregular singularity}
 \label{fig:3pointfusionchannel}
\end{figure}

\begin{figure}
    \centering
\includegraphics[scale=0.5]{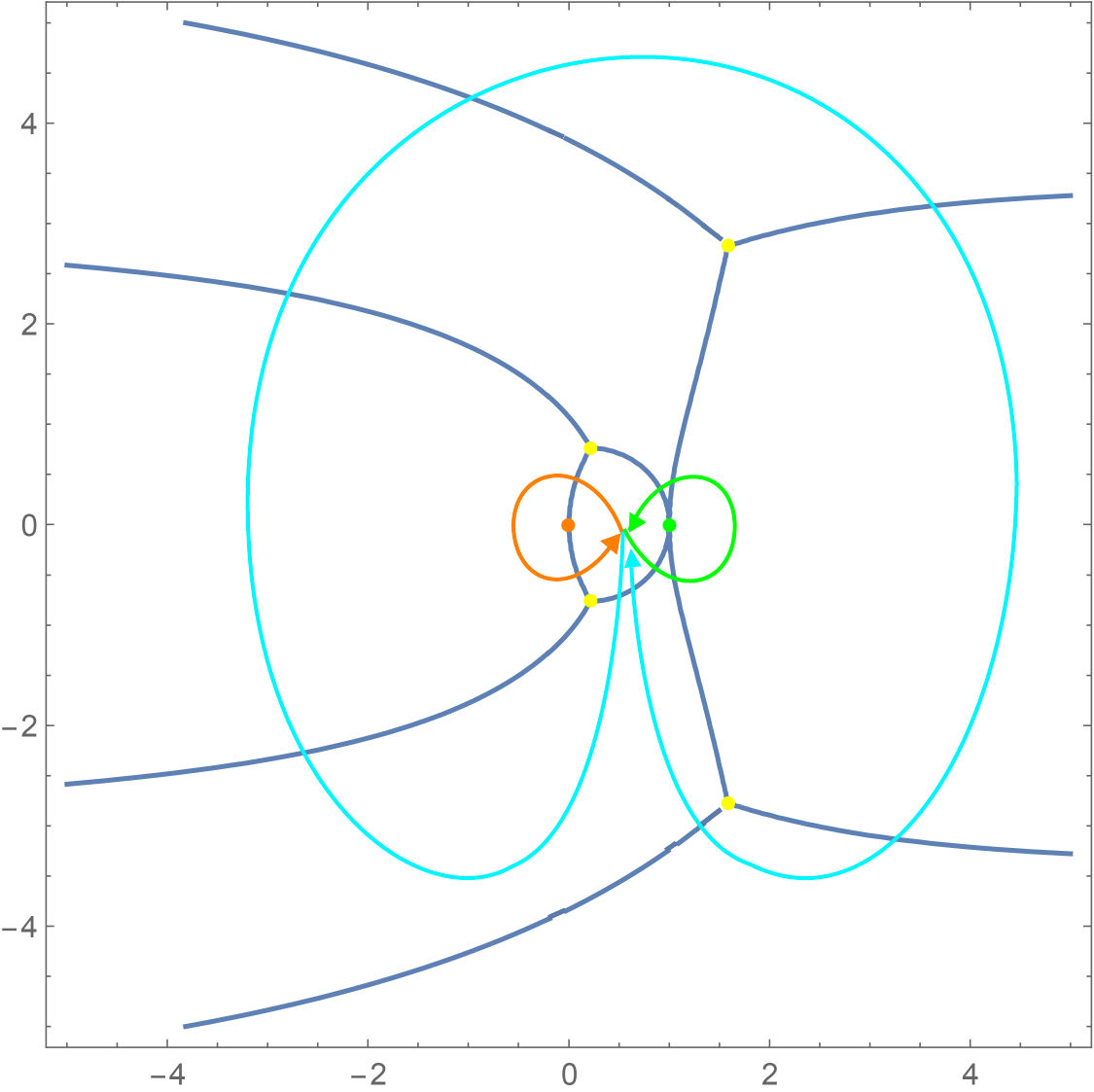}
    \caption{The Stokes lines on the $z$ plane.}
    \label{fig:higherstokes}
\end{figure}

Now we move to 3-point functions with all charges $k_a=1$. The relevant fusion rule is the same as in \cref{eq:fusionrule}, and the process of fusion is represented in \cref{fig:3pointfusionchannel}. The momentum of $(\Psi_1,\Psi_2)$ may be $(0,\frac{-1}{2b}),(\frac{-1}{b},\frac{-1}{2b}),(\frac{-1}{b},\frac{-3}{2b})$. Thus we expect a conformal block space of dimension $3$.
In this case there are no known special functions to compare with, and we will omit many of the details of computation since they are analogous to the case of Bessel functions considered before. We will list some of the intermediate results in the appendix.
The integral we want to study is 
\begin{equation}
\label{eq:3pointint}
\begin{aligned}
\mathcal{I}&=\oint dw[(z_1-z_2)(z_2-z_3)(z_1-z_3)]^{-1/(2b^2)}[(w-z_1)(w-z_2)(w-z_3)]^{1/b^2}\\
&\qquad\times\exp[\frac{\Lambda}{b^2}(z_1/2+z_2/2+z_3/2-w)]
\\
&=\oint dw\exp[\mathcal{W}/b^2].
\end{aligned}
\end{equation}
Sending $z_1\rightarrow 0,\; z_2\rightarrow 1,\; z_3\rightarrow z$,
\begin{equation}
	\begin{aligned}
		\mathcal{I}&=e^{\frac{\Lambda}{2b^2}(1+z)}[z(1-z)]^{-1/(2b^2)}\oint dw[w(w-1)(w-z)]^{1/b^2}\exp[-\frac{\Lambda}{b^2}w].
	\end{aligned}
\end{equation}
In other words, 
\begin{equation}
    \mathcal{W}=-\frac12\log z-\frac{1}{2}\log(z-1)+\log(w-z)+\log w+\log(w-1)+\Lambda(\frac{z}2-w).
\end{equation}
For simplicity we can take $\Lambda=1$.
This function has $3$ critical points, so we have $3$ corresponding thimbles. We can also infer from the expression that $\mathcal{I}$ has three singular points, namely $0,1, \infty$. The formal monodromy matrices of $0,1$ can be obtained by series expansion:
\begin{equation}
\hat{M}_0=\begin{pmatrix}
    q^{-3/2}&0&0\\
    0&q^{1/2}&0\\
    0&0&q^{1/2}
\end{pmatrix}
\quad \hat{M}_1=\begin{pmatrix}
    q^{1/2}&0&0\\
    0&q^{-3/2}&0\\
    0&0&q^{1/2}
\end{pmatrix}.
\end{equation}
The pattern of Stokes lines is shown in \cref{fig:higherstokes}. The position of $z=0$ and $z=1$ can be read from the scale. We will first work in the small bounded Stokes sector between these two points. 
In this sector, the thimbles are $\mathcal{J}_1=0\rightarrow z$, $\mathcal{J}_2=z\rightarrow 1$ and $\mathcal{J}_3=1\rightarrow \infty$. Then the monodromy matrices can be written in this basis as
\begin{equation}
    M_0=\begin{pmatrix} q^{-3/2}&0&0\\
    q^{1/2}-q^{-3/2}&q^{1/2}&0\\
    0&0&q^{1/2}
    \end{pmatrix}
    ,\quad
    M_1=\begin{pmatrix}
        q^{1/2}&q^{1/2}-q^{-3/2}&0\\
        0&q^{-3/2}&0\\
        0&0&q^{1/2}
    \end{pmatrix}.
\end{equation}
The paths of the analytic continuation is marked by lines with an arrow in \cref{fig:higherstokes} in the corresponding color.
To get the monodromy around infinity, we move to the region just below the center one for convenience. In this region, the thimble basis are
$0\rightarrow 1$, $z\rightarrow 1$, and $z\rightarrow \infty$.
In this basis, the monodromy around $\infty$ is
\begin{equation}
    M_\infty'=\begin{pmatrix}
        q^{-1}&0&0\\
        q^{-1}-q&q^3&0\\
        0&0&q^{-1}
    \end{pmatrix},
\end{equation}
marked by the big blue path.
We can use the Stokes matrices to change the basis back to the center region (see the appendix for the notations);
\begin{equation} \label{eq:Minf}
    M_\infty=(S_1^1)^{-1}M_\infty'S_1^1=\begin{pmatrix}
        q&q-q^3&0\\
        q^{-1}-q&q^{-1}-q+q^3&0\\
        0&0&q^{-1}
    \end{pmatrix}.
\end{equation}
From this we can verify that 
\begin{align}
    M_\infty M_1 M_0=I.
\end{align}
This can be readily seen from the pattern of the paths and the relations of generators in the 
homotopy group. Note that by diagonalizing equation \eqref{eq:Minf}, we can extract the ``F''-matrix corresponding to attaching the operator at position $z$ to the irregular line $\alpha$. However, due to the nonsimple nature of $\alpha$, there is no well-defined fusion rule between these two different operator types. Still our procedure shows, that this ``attaching'' procedure works and that one can perform a basis change and a subsequent local braiding move in this situation. 

\section{Braid group representations}
\label{sec:braiding}
The conformal blocks form also representations of the braid group. We would like to first consider the braid group with two generators $\mathcal{B}=\{\sigma_1,\sigma_2\}$. Our starting point is the integration \eqref{eq:3pointint}.
Contrary to the monodromy, here we cannot fix any of the points. We must allow all the $3$ points to move. Assume that we are in a region where thimbles are $\mathcal{J}_1=z_1\rightarrow z_2$, $\mathcal{J}_2=z_2\rightarrow z_3$, $\mathcal{J}_3=z_3\rightarrow \infty$. Let us first compute the action of $\sigma_1$ where we braid $z_1$ and $z_2$. To do this, we again need to draw the thimbles. 

The exchange of $z_1$ and $z_2$ will give the integration an overall phase $q^{1/4}$. There are two ways to ``exchange'' $2$ points. One way is to rotate them around their mid point, and the other is to fix one point while moving another point around the fixed one by $\pi$ angle. We will do both of them in the following. 

\begin{figure}
    \centering
    \includegraphics[scale=0.3,trim={0.4cm 9.5cm 0.5cm 5cm},clip]{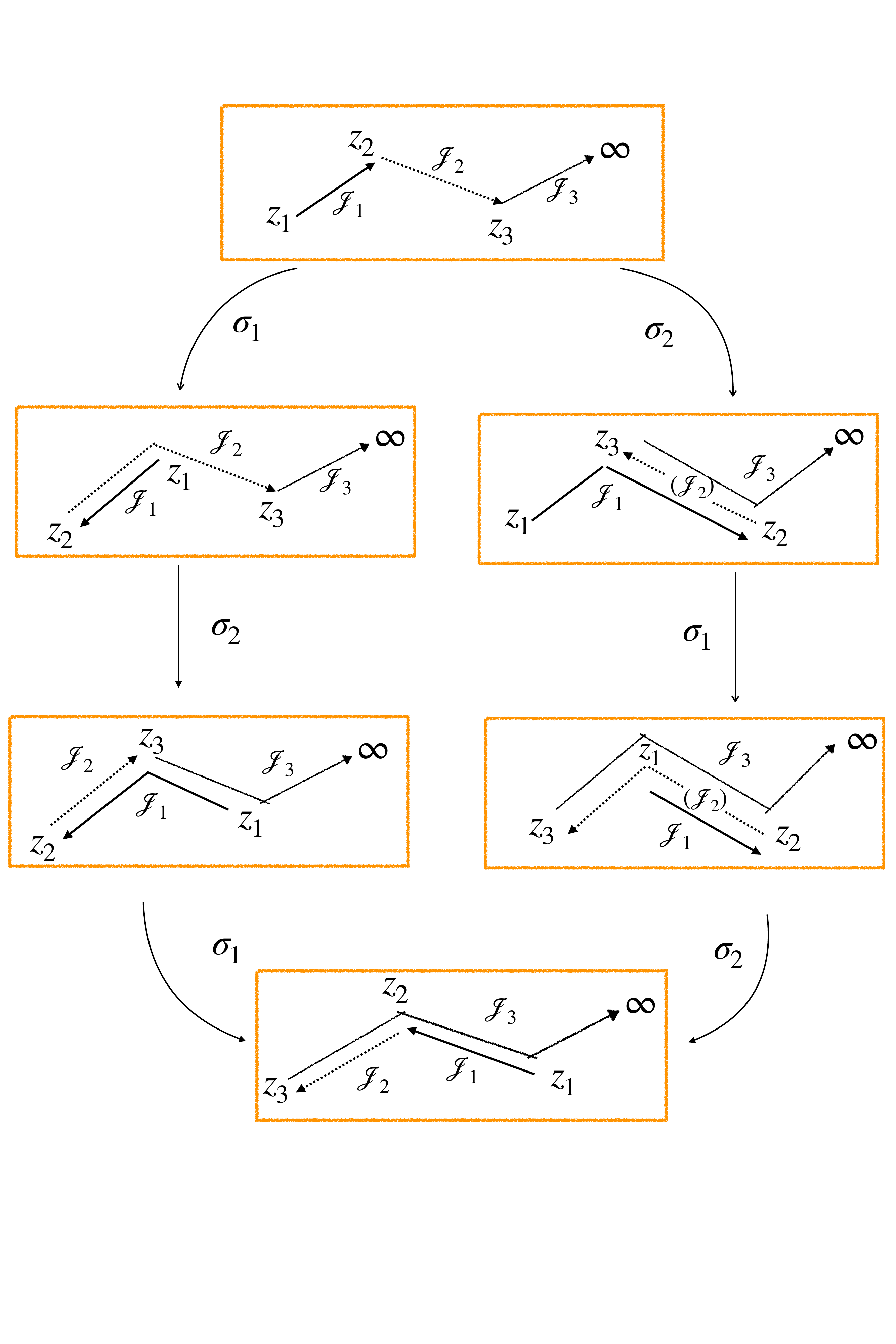}
    \caption{The braiding of conformal blocks. Note that $\sigma_1\sigma_2\sigma_1=\sigma_2\sigma_1\sigma_2$.}
    \label{fig:Braiding big}
\end{figure}

In the first ``rotating around the midpoint'' approach, a single action of $\sigma_1$ can be represented as
\begin{equation}
    \sigma_1[\begin{pmatrix}
         \mathcal{J}_1\\
         \mathcal{J}_2\\
         \mathcal{J}_3
     \end{pmatrix}]\doteq\begin{pmatrix}
        -q^{1/4}&0&0\\
        q^{1/4}&q^{1/4}&0\\
        0&0&q^{1/4}
    \end{pmatrix}
    \begin{pmatrix}
         \mathcal{J}_1\\
         \mathcal{J}_2\\
         \mathcal{J}_3
     \end{pmatrix}.\end{equation}
This can be seen from the left upper part of \cref{fig:Braiding big}. Note that $\mathcal{J}_1$ have changed its direction, and $[(w-z_1)(w-z_2)(w-z_3)]^{1/b^2}$ doesn't change its value if we swap any two of $z_i$s. Similarly the $\sigma_2$ can be represented in this basis as
\begin{equation}
    \sigma_2\doteq=\begin{pmatrix}
        q^{1/4}&q^{1/4}&0\\
        0&-q^{1/4}&0\\
        0&q^{1/4}&q^{1/4}
    \end{pmatrix}.
\end{equation}
An successive operation of $\sigma$'s can be represented as multiplications of the above matrices. Here one should be careful that the matrix of the later operation should multiply from the right side because we use a convention that basis (instead of coefficients) are column vectors.
Fig.\ref{fig:Braiding big} is a pictorial proof to the braiding relation $\sigma_1\sigma_2\sigma_1=\sigma_2\sigma_1\sigma_2$. 
The net effect of both two successive operations can be written as the following matrix:
\begin{equation}
    \begin{pmatrix}
        0&-q^{3/4}&0\\
        -q^{3/4}&0&0\\
        q^{3/4}&q^{3/4}&q^{3/4}
    \end{pmatrix}.
\end{equation}

\subsection{The second approach}
\begin{figure}
    \centering
    \includegraphics[scale=0.25,trim={13cm 5cm 15cm 3cm},clip]{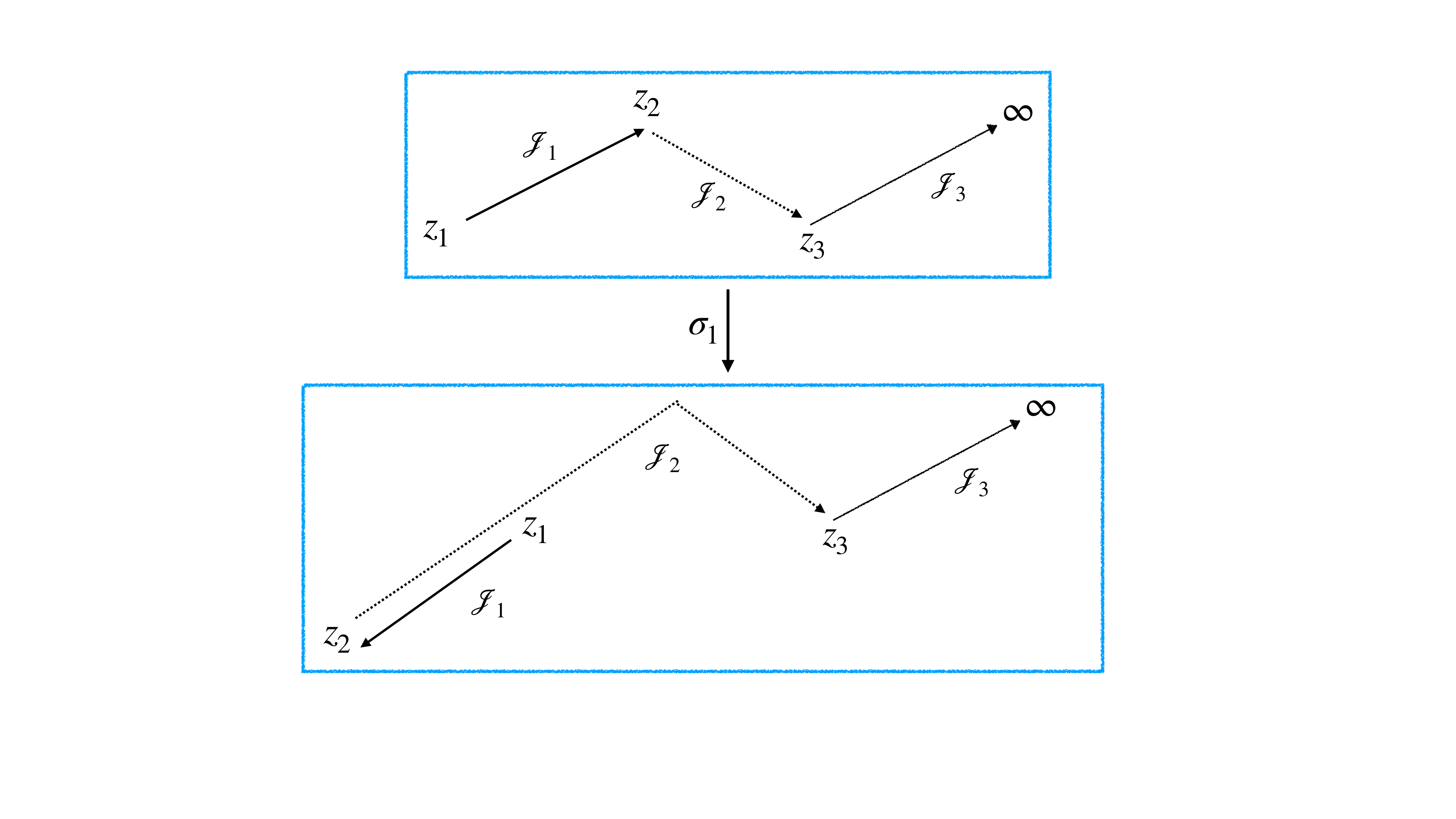}
    \caption{The action of $\sigma_1$}
    \label{fig:Small braiding}
\end{figure}

We would like to write down the action of single $\sigma_1$ as an example, see \cref{fig:Small braiding}: 
\begin{equation}
     \begin{pmatrix}
         \mathcal{J}_1'\\
         \mathcal{J}_2'\\
         \mathcal{J}_3'
     \end{pmatrix}=\begin{pmatrix}
        q^{-3/4}&0&0\\
        q^{1/4}-q^{-3/4}&q^{1/4}&0\\
        0&0&q^{1/4}
    \end{pmatrix}
    \begin{pmatrix}
         \mathcal{J}_1\\
         \mathcal{J}_2\\
         \mathcal{J}_3
     \end{pmatrix}.
\end{equation}

We can do a similar analysis to more complicated braiding moves. The pictorial representations are essentially the same as  in approach one, so we omit them. Instead, we write down the net effect of $\sigma_1\sigma_2\sigma_1$ to verify the braiding relation. This amounts to a succession of $3$ matrices:
\begin{equation}
\begin{pmatrix}
        q^{1/4}&0&0\\
        0&q^{-3/4}&0\\
        0&q^{1/4}-q^{-3/4}&q^{1/4}
    \end{pmatrix}
    \begin{pmatrix}
        q^{1/4}&0&0\\
        q^{-3/4}-q^{1/4}&q^{-3/4}&0\\
        q^{1/4}-q^{-3/4}&q^{1/4}-q^{-3/4}&q^{1/4}
    \end{pmatrix}
     \begin{pmatrix}
        q^{-3/4}&0&0\\
        q^{1/4}-q^{-3/4}&q^{1/4}&0\\
        0&0&q^{1/4}
    \end{pmatrix}.
\end{equation}
The action of $\sigma_2\sigma_1\sigma_2$ amounts to 
\begin{equation}
    \begin{pmatrix}
        q^{-3/4}&0&0\\
        q^{1/4}-q^{-3/4}&q^{1/4}&0\\
        0&0&q^{1/4}
    \end{pmatrix}
    \begin{pmatrix}
        q^{1/4}&0&0\\
        q^{-3/4}-q^{1/4}&q^{-3/4}&0\\
        q^{1/4}-q^{-3/4}&q^{1/4}-q^{-3/4}&q^{1/4}
    \end{pmatrix}
    \begin{pmatrix}
        q^{1/4}&0&0\\
        0&q^{-3/4}&0\\
        0&q^{1/4}-q^{-3/4}&q^{1/4}
    \end{pmatrix}.
\end{equation}
They both multiply to 
\begin{equation}
    \begin{pmatrix}
        q^{-1/4}&0&0\\
        0&q^{-5/4}&0\\
        q^{3/4}-q^{-1/4}&q^{3/4}-q^{-5/4}&q^{3/4}
    \end{pmatrix}.
\end{equation}
Thus we can see that the braiding relation $\sigma_1\sigma_2\sigma_1-\sigma_2\sigma_1\sigma_2=0$ holds.

 \section{Conclusion}
 \label{sec:conclusions}
In this paper we have studied the monodromy and braid group representations of conformal blocks with irregular operators in 2d Liouville theory using the Stokes phenomenon. The method can be generalized to braid group $\mathcal{B}_n$ of arbitrary rank $n$. One of the applications we have in mind is the description of modular tensor categories arising from taking the Liouville parameter $b^2$ to be rational as outlined in the introduction. We find that taking irregular operators into account leads to nonsimple objects in the tensor category which makes their braiding and fusion rules more interesting for future studies. 

As is well known, the study of CFT conformal blocks with primary and degenerate operators leads to Fuchsian differential equations whose solutions are often described in terms of Hypergeometric functions \cite{Belavin:1984vu}. The corresponding higher order differential equations can be recast in terms of vectorial first order ODE's of which the Knizhnik-Zamolodchikov (KZ) equations are the most well-known example \cite{KNIZHNIK198483}. Similarly, it is expected that taking irregular operators into account leads to modifications of KZ-equations along the lines of \cite{Resh-KZ,Feigin_2010a,Feigin_2010b,XuQG}. For future purposes, it would be interesting to derive similar differential equations satisfied by the conformal blocks studied in the current paper. One interesting application of such equations would be the explicit solution of conformal blocks as a series expansion in different Stokes sectors. As such series expansions would be asymptotic, there is the immediate question of their relation to quantum periods, resurgence and difference equations \cite{GuMarino}. 

Yet from another perspective, the regular and irregular operators extend, via their braiding relations, to line defects in the corresponding 3d TQFTs. Here the irregular operators will give rise to new types of one-form symmetries whose study is expected to be fruitful. Moreover, pushing such 3d line defects to the boundary of spacetime gives rise to line defects in the corresponding boundary CFT where many questions concerning their quantum dimension and many other properties relevant for fusion categories can be asked  \cite{Chang:2018iay}.
 
 \section*{Acknowledgements}
We would like to thank Sergei Gukov, Bo Lin, Yihua Liu, Nicolai Reshetikhin and Youran Sun  for valuable discussions. This work is supported by the National Thousand Young Talents grant of China and the NSFC grant 12250610187.

\appendix

\section{Stokes matrices in 3 point conformal blocks}
In \cref{fig:higherstokes}, the orange path circling around $z=0$ intersects Stokes lines twice. Each intersection corresponds to a Stokes matrix. We label them in order as: 
\begin{equation}
    S_0^1=\begin{pmatrix}
        1&0&0\\
        1&1&0\\
        0&0&1      \end{pmatrix},
      S_0^2=  \begin{pmatrix}
            1&0&0\\
            -q^2&1&0\\
            0&0&1
        \end{pmatrix}.
\end{equation}
Similarly, along the green path circling $z=1$ there are four Stokes matrices:
\begin{equation}
    S_1^1=\begin{pmatrix}
       1&1&0\\
       0&1&0\\
       0&0&1
    \end{pmatrix},
    S_1^2=\begin{pmatrix}
        1&0&0\\
        0&1&0\\
        0&1&1
    \end{pmatrix},
    S_1^3=\begin{pmatrix}
        1&0&0\\
        0&1&0\\
        0&-1&1
    \end{pmatrix},
   S_1^4=\begin{pmatrix}
       1&-q^2&0\\
       0&1&0\\
       0&0&1
    \end{pmatrix}.
\end{equation}
If we analytically continue along the long blue path, $z$ will pass $8$ Stokes lines. But the first one and the last one correspond to $S_1^1$. We label the left $6$ Stokes matrices as
\begin{equation}
\begin{aligned}
	S_{\infty}^{1}&=\begin{pmatrix}
	1&0&0\\
	-1&1&0\\
	0&0&1
\end{pmatrix},
	S_{\infty}^{2}=\begin{pmatrix}
	1&0&0\\
	q^2&1&0\\
	0&0&1
\end{pmatrix},
	S_{\infty}^{3}=\begin{pmatrix}
	1&0&0\\
	0&1&1\\
	0&0&1
\end{pmatrix},
\\
S_{\infty}^{4}&=
\begin{pmatrix}
	1&0&0\\
	0&1&0\\
	0&-1&1
\end{pmatrix},
S_{\infty}^{5}=
\begin{pmatrix}
		1&0&0\\
		0&1&0\\
		0&1&1
\end{pmatrix},
S_{\infty}^{6}=
\begin{pmatrix}
	1&0&0\\
	0&1&-1\\
	0&0&1
\end{pmatrix}.
\end{aligned}
\end{equation}

\bibliographystyle{JHEP}     
{\small{\bibliography{main}}}
\end{document}